\documentclass[%
reprint,
superscriptaddress,
frontmatterverbose,
showpacs,preprintnumbers,
 amsmath,amssymb,
 aps,pra,floatfix, 
citeautoscript,
]{revtex4-2}

\usepackage{amsthm}
\theoremstyle{plain}
\newtheorem{theorem}{Theorem}
\newtheorem{proposition}[theorem]{Proposition}
\newtheorem{lemma}[theorem]{Lemma}
\newtheorem{corollary}[theorem]{Corollary}
\usepackage{algorithm}
\usepackage{algpseudocode}
\usepackage{mdframed}
\usepackage{mathtools}
\usepackage{times}

\usepackage[pdftex]{graphicx}

\usepackage{mathrsfs}
\usepackage[linkcolor=teal,colorlinks=true,citecolor=teal,filecolor=teal]{hyperref}
\usepackage{amsmath}
\usepackage[english]{babel}
\usepackage{booktabs}
\usepackage{amssymb}
\usepackage{type1cm}
\usepackage{braket}
\usepackage{tikz}
\usepackage[most]{tcolorbox}
\usepackage{mathtools}
\usepackage{pgfplots}
\usepackage{lipsum}
\usepackage{xcolor}
\usepackage{graphicx}
\usepackage{subcaption}
\usepackage{dcolumn}
\usepackage{bm}
\usepackage{physics}
\usepackage[utf8]{inputenc}
\usepackage{multirow}
\usepackage{tikz}

\usepackage{cleveref}

\crefname{equation}{Eq.}{Eqs.}
\Crefname{equation}{Equation}{Equations}
\crefname{figure}{Fig.}{Figs.}
\Crefname{figure}{Figure}{Figures}
\crefname{figure}{Fig.}{Figs.}
\Crefname{figure}{Figure}{Figures}
\crefname{section}{Supplemental Material Section}{Supplemental Material Sections}
\Crefname{section}{Supplemental Material Section}{Supplemental Material Sections}
\crefname{appendix}{Appendix}{Appendices}
\Crefname{appendix}{Appendix}{Appendices}
\crefname{table}{Table}{Tables}
\Crefname{table}{Table}{Tables}

\usepackage{enumitem,amssymb}
\newlist{todolist}{itemize}{2}
\setlist[todolist]{label=$\square$}
\usepackage{pifont}

\usepackage{etoolbox}
\pgfplotsset{compat=1.18} 
\begin{document}

\title{Tomography of approximately Gaussian mixed states}

\author{S. Nibedita Swain}
\email{swain.snibedita@gmail.com}
\affiliation{Centre for Quantum Software and Information, University of Technology Sydney, Sydney, New South Wales 2007, Australia}
\affiliation{School of Mathematical and Physical Sciences,
University of Technology Sydney, Ultimo, NSW 2007, Australia}

\affiliation{Sydney Quantum Academy, Sydney, NSW 2000, Australia}
\affiliation{Centre for Quantum Computation and Communication Technology, School of Mathematics
and Physics, University of Queensland, Brisbane, Queensland 4072, Australia}

\author{Timothy C. Ralph}
\email{ralph@physics.uq.edu.au}
\affiliation{Centre for Quantum Computation and Communication Technology, School of Mathematics
and Physics, University of Queensland, Brisbane, Queensland 4072, Australia}

\author{Jens Eisert}
\email{jense@zedat.fu-berlin.de}
\affiliation{Dahlem Center for Complex Quantum Systems, Freie Universität Berlin, 14195 Berlin, Germany}
\affiliation{Helmholtz-Zentrum Berlin für Materialien und Energie, 14109 Berlin, Germany}

\author{Nathan Walk}
\email{nathan.walk@gmail.com}
\affiliation{Dahlem Center for Complex Quantum Systems, Freie Universität Berlin, 14195 Berlin, Germany}

\begin{abstract}
Moment-based tomography efficiently reconstructs Gaussian states,
but stochastic phase noise generally produces mixed, non-Gaussian
outputs. For approximately Gaussian outputs, we formulate reconstruction
as a problem in quantum learning theory under model misspecification.
Our trace-distance guarantee separates the intrinsic Gaussian approximation
error from statistical errors in estimating the first moments and
covariance matrix. Using recent Gaussian trace-distance bounds, we
obtain linear dependence of the statistical contribution on moment
errors when the covariance inverses remain bounded, improving on
the square-root dependence on covariance error in earlier guarantees.
A Fock-space truncation procedure provides certified, convergent
lower and upper bounds on the intrinsic error. Under an explicit sub-Gaussian condition on the heterodyne outcomes, the sufficient sample count
scales inversely with the square of the allowed statistical error 
remaining after accounting for the approximation error. We verify the statistical condition for stochastic phase rotations and heralded unitary-averaged states with bounded input squeezing. We apply
the framework to probabilistic unitary averaging for phase noise
reduction, quantifying reconstruction error and accounting for
heralding overhead in the total number of estimation attempts.
\end{abstract}

\date{\today}
\frenchspacing
\maketitle

\section{Introduction}

Gaussian states constitute one of the most important and experimentally accessible classes of \emph{continuous-variable} (CV) quantum states~\cite{eisertplenio2003basics,weedbrook2012gaussian,BAC19}, underpinning applications in quantum communication~\cite{gisin2007quantum}, sensing and metrology~\cite{polino2020photonic}, and quantum teleportation~\cite{furusawa1998unconditional, pirandola2015advances, suzuki2026ultrafast}. Their defining advantage is that they admit a compact description in terms of their first moments and covariance matrices~\cite{serafini2023quantum}. Consequently, Gaussian state tomography is considerably more tractable than the reconstruction of an arbitrary CV state: homodyne or heterodyne
measurements can be used to estimate the relevant moments without 
reconstructing an infinite-dimensional density matrix~\cite{vrehavcek2015surmounting, mele2025learning}. This experimental accessibility has become increasingly important with the rapid development of quantum technologies in CV photonic platforms~\cite{madsen2022quantum, liu2026gaussian}, creating a growing demand for tomography protocols that are both experimentally feasible and supported by rigorous quantitative recovery guarantees~\cite{anshu2024survey}. Although substantial progress has been made in finite-dimensional systems~\cite{cramer2010efficient, lanyon2017efficient, hangleiter2024bell, arunachalam2022optimal, austin2025efficiently, landau2025learning}, comparable results for CV systems have emerged only recently because their infinite-dimensional Hilbert spaces introduce qualitatively new conceptual and technical challenges~\cite{mele2025learning, bittel2025optimal, fanizza2024efficient, holevo2024estimates, gandhari2024precision, upreti2024efficient,zhao2025complexity, arzani2025effective}. Nevertheless, physically realisable CV states are not arbitrary infinite-dimensional states: finite experimental resources impose constraints such as finite mean energy or mean total photon number~\cite{serafini2023quantum}, providing essential structure for the development of practical tomography guarantees.

The rigorous study of Gaussian state tomography has advanced rapidly in recent years.
A recent review provides an overview of bosonic quantum learning, including tomography, non-Gaussianity, and Gaussianity testing~\cite{mele2026bosonicreview}. Although tomography of general energy-constrained CV states is inefficient, assuming Gaussianity enables efficient reconstruction from first and second moments. Trace-distance guarantees (upper bounds on the trace distance between the reconstructed and true states) have been derived from moment estimation errors, with extensions to weak non-Gaussian perturbations, quantified by the relative entropy of non-Gaussianity, and to states generated by 
a 
small number of local non-Gaussian evolutions~\cite{mele2025learning}. Subsequent work established optimal trace-distance bounds with linear dependence on displacement and covariance estimation errors~\cite{bittel2025optimal}. Further developments improved the guarantees for heterodyne tomography and introduced adaptive protocols whose sample complexity is essentially independent of the state energy~\cite{bittel2025energy}. The optimal sample complexity scaling has also been determined for learning pure and mixed bosonic and fermionic Gaussian states without an energy constraint~\cite{chen2026optimal}.

On the experimental side, moment-based multiparameter estimation has enabled the squeezing, antisqueezing, and squeezing angle of squeezed vacuum states to be determined from homodyne and double-homodyne measurements at the Cram\'er--Rao limit~\cite{tripier2025optimal}. Collectively, these results establish a mature framework for tomography under a Gaussian promise, together with partial extensions to restricted classes of non-Gaussian states.

Despite this progress, the Gaussian promise is often violated in realistic experiments. A particularly important example is stochastic phase noise. Conditioned on a fixed phase realisation, the evolution is Gaussian and therefore maps a Gaussian input to a Gaussian output. When the phase fluctuates between experimental runs and remains unobserved, however, the detector sees an ensemble average over differently transformed Gaussian states. This mixture is generally mixed and non-Gaussian, although it may remain close to Gaussian for weak phase fluctuations. The same mechanism persists after probabilistic phase noise reduction protocols~\cite{swain2024improving, swain2025noise, swain2026enhancing}: postselection may improve squeezing, purity, and entanglement, yet averaging over the residual unobserved phases generally leaves a non-Gaussian accepted ensemble, whose deviation from Gaussianity must be quantified.

For moment-based Gaussian tomography, this creates a genuine model-mismatch problem. The measured first and second moments specify a unique Gaussian state with the same first moments and covariance matrix, but they do not determine the full density matrix. An operational trace-distance guarantee must therefore separate two contributions: the intrinsic error from approximating the density matrix by the Gaussian state with the same first moments and covariance matrix, and the statistical error arising from finite-sample estimation of the moments. Increasing the number of measurements can suppress the statistical error, whereas the intrinsic contribution is the asymptotic approximation error of a consistent Gaussian reconstruction based on matching the first two moments.

Quantum learning theory studies state reconstruction within a chosen family of states, called a hypothesis class, at a specified accuracy and confidence~\cite{arunachalamdewolf2017survey}. Here, this family consists of Gaussian states, although stochastic phase noise may make the target state non-Gaussian. The model is then said to be misspecified. Our estimator is proper because it always returns a physical Gaussian state. We bound its trace-distance error relative to the actual target by two contributions such as an approximation error and a statistical error from estimating moments using finitely many measurements. The approximation error is the distance between the target and the Gaussian state with the same first moments and covariance matrix. This Gaussian reference need not be the closest Gaussian state in trace distance. Increasing the sample size reduces the statistical error, while describing the target's non-Gaussian features requires a more general reconstruction model.

Here we present a tomography framework for approximately Gaussian mixed
states. Our reconstruction bound separates the intrinsic error of
approximating the output by the Gaussian state with the same first
moments and covariance matrix from the statistical error in estimating
those moments. Fock-space truncation provides convergent lower and
upper bounds on the approximation error.  The inverse square dependence of our sample
bound on the allowed statistical error follows by combining sub-Gaussian moment concentration with
the more recent Gaussian bounds of
Refs.~\cite{bittel2025optimal,bittel2025energy}. Finally, we apply the framework to the unitary
averaging protocol of Ref.~\cite{swain2024improving}, quantify the
reconstruction error of its heralded output, and account for the
success probability when estimating the total number of experimental
attempts.

\section{Bound for the approximately Gaussian mixed state}

We first specify the target state, the Gaussian reconstruction rule, and the loss used to assess it.
Theorem~\ref{theorem1} converts moment errors into a deterministic reconstruction bound, while Theorem~\ref{theorem2} turns concentration of heterodyne data into a finite-sample learning guarantee.
The distinction between these steps makes the statistical assumptions explicit.


\subsection{Setting considered}
Gaussian states and Gaussian operations constitute an important resource for continuous-variable quantum communication, sensing, and information processing. Their practical performance, however, can be degraded by stochastic phase fluctuations arising during state preparation, transmission, or interferometric processing. Although a fixed phase shift is a Gaussian unitary and therefore preserves Gaussianity, an unobserved phase that fluctuates between experimental runs produces an ensemble average over differently rotated Gaussian states. Such an average generally reduces purity, squeezing, and entanglement and, crucially, need not itself be Gaussian.

The problem of interest is therefore not limited to characterising the degradation caused by phase noise. It also arises when certifying protocols designed to suppress that noise. For example, Ref.~\cite{swain2024improving} proposed a passive linear-optical unitary-averaging (UA) protocol in which an unknown Gaussian signal is coherently distributed across several independently noisy channel uses, recombined, and kept following vacuum heralding (more details in Appendix~\ref{Unitary Averaging in a nutshell}) and UA protocol also discussed for discrete variable cases in Refs.~\cite{marshman2018passive, marshman2024unitary,singh2026coherently}. In Ref.~\cite{swain2024improving}, the protocol reduces effective phase fluctuations and improves the transmitted purity, squeezing, and entanglement. Nevertheless, the residual phases remain stochastic. Consequently, even when the mitigation protocol is successful, averaging the accepted output over the unobserved phase realisations generally produces a mixed, generally weakly non-Gaussian state. Whether a Gaussian reconstruction meets a chosen accuracy is determined by the approximation error certificate below.
This setting motivates a tomography question: how accurately can a Gaussian hypothesis learned from the first two moments represent the accepted, generally non-Gaussian ensemble?
A reconstruction guarantee must distinguish the intrinsic error of that hypothesis from the statistical error caused by estimating its moments from finitely many accepted runs.

Let
\begin{equation}
\rho_{\mathrm{in}}
=
\rho_{\mathrm{G}}
\bigl(\Sigma_{\mathrm{in}},m_{\mathrm{in}}\bigr)
\end{equation}
be an $n$-mode Gaussian input state. We first consider a stochastic phase noise channel of the form
\begin{equation}
\mathcal{P}_{\nu}(\rho)
\coloneqq
\int
p_{\nu}(\phi)\,
U_{\phi}\rho U_{\phi}^{\dagger}\,
d\phi,
\end{equation}
where $\phi$ denotes the relevant phase variable, or collection of phase variables, $p_{\nu}(\phi)$ is its probability distribution parametrised by the noise strength $\nu$, and $U_{\phi}$ is the corresponding Gaussian phase-shift unitary. The ensemble-averaged output state is
given by
\begin{equation}
\varrho_{\nu}
\coloneqq
\mathcal{P}_{\nu}(\rho_{\mathrm{in}})
=
\int
p_{\nu}(\phi)\,
U_{\phi}\rho_{\mathrm{in}}U_{\phi}^{\dagger}\,
d\phi.
\label{Eqphasenoisy}
\end{equation}

For each fixed phase realisation $\phi$, the state
$U_{\phi}\rho_{\mathrm{in}}U_{\phi}^{\dagger}$ is Gaussian. Gaussian states are not, however, closed under arbitrary convex mixtures, and hence \(\varrho_{\nu}\) is generally a mixed non-Gaussian state. When the phase distribution becomes sufficiently concentrated, the output keeps an approximately Gaussian character. In particular, if $p_{\nu}(\phi)$ converges to a deterministic phase distribution as $\nu\to0$, then $\varrho_{\nu}$ converges to a Gaussian state and the Gaussian approximation becomes exact. This is the weak-noise regime underlying the moment-based analysis of Ref.~\cite{swain2024improving}.

More generally, the deterministic reconstruction bound and the truncation certificate apply to any state with finite second moments. The sample bound additionally requires the heterodyne tail condition in Theorem~\ref{theorem2}. In that setting, $\varrho_{\nu}$ is understood as the normalised ensemble-averaged state of the accepted experimental runs, and $\nu$ characterises the residual phase noise. 

\subsection{Preliminaries}

Let $R=(R_1,\ldots,R_{2n})^{\mathrm{T}}$ 
denote the vector of canonical quadrature operators, satisfying the \emph{canonical commutation relations} (CCR) giving rise to the symplectic form $\Omega$.
We assume finite first and second moments throughout. The first moment vector $m\in\mathbb{R}^{2n}$ and symmetric 
covariance matrix $\Sigma\in\mathbb{R}^{2n\times2n}$ of $\varrho_{\nu}$ are defined by
\begin{align}
m_j
&\coloneqq
\operatorname{Tr}\!\left(\varrho_{\nu}R_j\right),\\
\Sigma_{jk}
&\coloneqq
\frac{1}{2}
\operatorname{Tr}\!\left[
\varrho_{\nu}
\left\{
R_j-m_j,\,
R_k-m_k
\right\}
\right].
\end{align}
With this convention, the vacuum covariance matrix is the identity. We denote by
$\rho_{\mathrm{G}}(\Sigma,m)$ the Gaussian state having the same first and second moments as $\varrho_{\nu}$.

The intrinsic error incurred by replacing the physical output with its moment-matched Gaussian reference state is quantified by
\begin{equation}
\delta_{\mathrm{NG}}(\nu)
\coloneqq
D_{\mathrm{tr}}\!\left(
\varrho_{\nu},
\rho_{\mathrm{G}}(\Sigma,m)
\right)
=
\frac{1}{2}
\left\|
\varrho_{\nu}
-
\rho_{\mathrm{G}}(\Sigma,m)
\right\|_1.
\label{non-gaussian-approximation-error}
\end{equation}
Thus, \(\delta_{\mathrm{NG}}(\nu)\) measures the information about the output that is not captured by its first two moments. In the weak-noise regime, this quantity is expected to be small, and it vanishes when the output becomes exactly Gaussian.

Suppose that measurement data provide estimators
$\widetilde{\Sigma}$ and $\widetilde{m}$ satisfying
\begin{equation}
\left\|
\widetilde{\Sigma}-\Sigma
\right\|_{\infty}
\leq
\varepsilon,
\:\:
\left\|
\widetilde{m}-m
\right\|_2
\leq
\varepsilon,
\label{moment-estimation-errors}
\end{equation}
where \(\|\cdot\|_{\infty}\) denotes the operator norm and
\(\|\cdot\|_2\) denotes the Euclidean norm. We assume that
\(\widetilde{\Sigma}\) is a valid quantum covariance matrix; namely, it is real and symmetric and satisfies the uncertainty relation
\begin{equation}
\widetilde{\Sigma}+i\Omega\succeq0.
\end{equation}
The reconstructed Gaussian state
\begin{equation}
\widetilde{\rho}_{\mathrm{G}}
\coloneqq
\rho_{\mathrm{G}}
\bigl(\widetilde{\Sigma},\widetilde{m}\bigr)
\end{equation}
is therefore well defined.

By the triangle inequality, the distance between the density matrix and the reconstructed Gaussian state satisfies
\begin{align}
D_{\mathrm{tr}}\!\left(
\varrho_{\nu},
\widetilde{\rho}_{\mathrm{G}}
\right)
\leq{}&
D_{\mathrm{tr}}\!\left(
\varrho_{\nu},
\rho_{\mathrm{G}}(\Sigma,m)
\right)
\nonumber\\
&+
D_{\mathrm{tr}}\!\left(
\rho_{\mathrm{G}}(\Sigma,m),
\rho_{\mathrm{G}}
\bigl(\widetilde{\Sigma},\widetilde{m}\bigr)
\right)
\nonumber\\
={}&
\delta_{\mathrm{NG}}(\nu)
+
D_{\mathrm{tr}}\!\left(
\rho_{\mathrm{G}}(\Sigma,m),
\rho_{\mathrm{G}}
\bigl(\widetilde{\Sigma},\widetilde{m}\bigr)
\right).
\end{align}
The first term is an intrinsic model-mismatch error: it remains even if the moments are known exactly. The second is an estimation error arising from finite-precision knowledge of the first two moments and can, in principle, be reduced by collecting additional measurement data.

Our objective is to derive an explicit upper bound on the second contribution in terms of $\Sigma,\widetilde{\Sigma},m$, and $\widetilde{m}$, and thereby obtain a quantitative reconstruction guarantee for a mixed output state, either before or after phase noise mitigation. We quantify the reconstruction error using the trace distance,
\begin{equation}
D_{\mathrm{tr}}(\rho,\sigma)
\coloneqq
\frac{1}{2}\|\rho-\sigma\|_1.
\end{equation}
Its operational meaning follows from the Holevo-Helstrom theorem
\cite{helstrom1969quantum,holevo1976investigations}: for two states with
equal prior probabilities, the optimal probability of distinguishing them
is
\begin{equation}
P_{\mathrm{succ}}^{\mathrm{opt}}
=
\frac{1}{2}
\left[1+D_{\mathrm{tr}}(\rho,\sigma)\right].
\end{equation}
Consequently, trace-distance bounds for Gaussian states provide an
operationally meaningful measure of the reconstruction error caused by
imperfectly estimated moments~\cite{bittel2025energy}.

\subsection{Main results}

The specific question addressed here is the following: \emph{how do the residual non-Gaussianity generated by stochastic phase fluctuations and the errors in estimating the first two moments jointly determine the trace-distance error of a reconstructed state, both for an unprotected noisy channel and for the output of a reduction protocol?}


\begin{theorem}(Trace-distance bound for an approximately Gaussian mixed state.)
\label{theorem1}
Under the setting and notation introduced above, let $\Sigma$ and $\widetilde{\Sigma}$ be physical covariance matrices and let $m,\widetilde m\in\mathbb R^{2n}$. Then the Gaussian reconstruction satisfies
\begin{align}
D_{\mathrm{tr}}\!\left(
\varrho_{\nu},
\widetilde{\rho}_{\mathrm{G}}
\right)
\leq{}&
\delta_{\mathrm{NG}}(\nu)
+
\frac{1+\sqrt{3}}{8}
\operatorname{Tr}\!\left[
\left(
\Sigma^{-1}
+
\widetilde{\Sigma}^{-1}
\right)
\left|
\widetilde{\Sigma}-\Sigma
\right|
\right]
\nonumber\\
&+
\frac{1}{2}
\left\|
\widetilde{\Sigma}^{-1/2}
\left(m-\widetilde{m}\right)
\right\|_{2}.
\label{main-trace-bound}
\end{align}
Here, $\left|\widetilde{\Sigma}-\Sigma\right|$ denotes the matrix
absolute value of the real symmetric matrix
$\widetilde{\Sigma}-\Sigma$. A proof is given in
Appendix~\ref{Proof of Theorem 1}.
\end{theorem}

In Appendix~\ref{Proof of Theorem 1}, the proof is organised into three
steps. First, we introduce the Gaussian reference state
$\rho_{\mathrm G}(\Sigma,m)$ and apply the triangle inequality to
separate the non-Gaussian approximation error
$\delta_{\mathrm{NG}}(\nu)$ from the Gaussian reconstruction error.
Introducing the intermediate state
$\rho_{\mathrm G}(\widetilde{\Sigma},m)$ then separates the latter
into covariance and displacement contributions.

Second, we bound the covariance contribution using the derivative
framework developed in Ref.~\cite{bittel2025optimal}, together with the
inverse-covariance derivative estimate established
in Ref.~\cite{bittel2025energy}. Unitary invariance under a common
displacement reduces this comparison to Gaussian states with zero
first moments. We then integrate the derivative estimate along the
straight-line path
$\Sigma_s=(1-s)\Sigma+s\widetilde{\Sigma}$, with $s\in[0,1]$.
Applying the operator convexity of matrix inversion yields the
endpoint weight
$\Sigma^{-1}+\widetilde{\Sigma}^{-1}$ 
and reproduces the covariance
term of the perturbation bound in Ref.~\cite{bittel2025energy}.

Third, we bound the displacement contribution by applying the
Fuchs--van de Graaf inequality to the fidelity between Gaussian
states with the common covariance matrix
$\widetilde{\Sigma}$~\cite{banchi2015quantum}. This gives the weighted
displacement term
$\frac{1}{2}
\|\widetilde{\Sigma}^{-1/2}(m-\widetilde m)\|_2$.
This step uses the mixed state fidelity directly:
\cite{bittel2025optimal} instead treats displacement errors through
a Gaussian noise channel and pure state overlaps, whereas
\cite{bittel2025energy} uses a first-moment derivative estimate.
Our theorem~1 therefore combines the existing
inverse covariance bound with a direct fidelity estimate and an
explicit additive contribution accounting for the non-Gaussian
approximation error.

We formulate the reconstruction as a learning algorithm that takes independent heterodyne measurement outcomes as input and returns a physical Gaussian state. The target accuracy is $\varepsilon_{\mathrm{tot}}$, with confidence $1-\delta$. To bound the required sample size, we assume a tail condition controlling the probability of unusually large outcomes under the actual measurement distribution, which may be non-Gaussian. The bounds $U$ and $B$, introduced below, specify the available information about the Gaussian approximation error and covariance, respectively. These bounds are assumed known before the sample size is chosen.


\begin{theorem}[Sample complexity for heterodyne reconstruction of an approximately Gaussian mixed state]
\label{theorem2}
Let $\varrho_\nu$ be an $n$-mode state with finite first and second moments $(m,\Sigma)$. A heterodyne outcome $X\in\mathbb R^{2n}$ is normalised so that
\begin{equation}
\mathbb E[X]=m,\: H \coloneqq \operatorname{Cov}(X)=\Sigma+I_{2n}.
\end{equation}
Assume that a known parameter $\kappa\geq1$ satisfies
\begin{equation}
\mathbb E\!\left[e^{u^{\mathrm T}(X-m)}\right]
\leq \exp\!\left(\frac{\kappa^2}{2}u^{\mathrm T}Hu\right)
\quad\text{for every }u\in\mathbb R^{2n}.
\label{eq:heterodyne-subgaussian}
\end{equation}
Let $X_1,\ldots,X_N$ be independent outcomes from $N$ copies of $\varrho_\nu$, and define
\begin{align}
\widehat m_H& \coloneqq \frac1N\sum_{j=1}^N X_j,\label{eq:empirical-mean}\\
\widehat H& \coloneqq \frac1N\sum_{j=1}^N
(X_j-\widehat m_H)(X_j-\widehat m_H)^{\mathrm T}.
\label{eq:empirical-covariance}
\end{align}
For $\delta\in(0,1)$, set
\begin{align}
q_\delta&:=2n\ln9+\ln(4/\delta),\qquad
E\coloneqq 2n+\operatorname{Tr}(\Sigma^{-1}),\label{eq:q-and-E}\\
\zeta_N&\coloneqq 24\kappa^2\left(\sqrt{\frac{q_\delta}{N}}+
\frac{q_\delta}{N}\right).
\label{eq:zeta-definition}
\end{align}
When $\zeta_N\leq1/2$, form the candidate moments
\begin{equation}
\widetilde m=\widehat m_H,\qquad
\widetilde\Sigma=\frac{\widehat H}{1-\zeta_N}-I_{2n}.
\label{eq:physical-estimator}
\end{equation}
If this candidate does not satisfy $\widetilde\Sigma+i\Omega\succeq0$, replace $(\widetilde\Sigma,\widetilde m)$ by $(I_{2n},0)$. If $\zeta_N>1/2$, also use this fixed vacuum output. Thus the estimator $\widetilde\rho_{\mathrm G}:=\rho_{\mathrm G}(\widetilde\Sigma,\widetilde m)$ is always a physical state.

Whenever $\zeta_N\leq1/2$, with probability at least $1-\delta$ the candidate in Eq.~\eqref{eq:physical-estimator} is physical and
\begin{align}
D_{\mathrm{tr}}(\varrho_\nu,\widetilde\rho_{\mathrm G})
\leq{}&\delta_{\mathrm{NG}}(\nu)
+\frac{1+\sqrt3}{2}\frac{\zeta_N}{1-\zeta_N}E\nonumber\\
&+\kappa\sqrt{\frac{q_\delta}{N}}
\sqrt{1+\|\Sigma^{-1}\|_\infty}.
\label{eq:finite-N-trace-bound}
\end{align}
In particular, suppose that fixed, known bounds $U$ and $B$ satisfy
\begin{equation}
\delta_{\mathrm{NG}}(\nu)\leq U<\varepsilon_{\mathrm{tot}}\leq1,
\qquad E\leq B.
\label{eq:known-bounds}
\end{equation}
There is a universal constant $C_*>0$ such that
\begin{equation}
N\geq C_*\,
\frac{\kappa^4 q_\delta B^2}{(\varepsilon_{\mathrm{tot}}-U)^2}
\label{eq:subgaussian-sample-bound}
\end{equation}
implies
\begin{equation}
\Pr\!\left[D_{\mathrm{tr}}(\varrho_\nu,\widetilde\rho_{\mathrm G})
\leq\varepsilon_{\mathrm{tot}}\right]\geq1-\delta.
\label{eq:tomography-success}
\end{equation}
One may take $C_*=2^{16}$; this constant is not optimised. Consequently, it is sufficient to use
\begin{equation}
N=\mathcal O\!\left(
\frac{\kappa^4[n+\ln(4/\delta)]B^2}
{(\varepsilon_{\mathrm{tot}}-U)^2}\right).
\label{sample-complexity-scaling}
\end{equation}
\end{theorem}

Trace distance bounds the difference between the probabilities predicted by the reconstructed Gaussian state and those of the actual state for any measurement event. The same total error bound therefore applies to measurements beyond the heterodyne measurements used for reconstruction. The confidence level describes how reliably this guarantee holds over repeated collections of measurement data. The target state is the fixed state obtained by averaging over the preparation noise. For a heralded preparation, it is the average state conditioned on acceptance.

Theorem~\ref{theorem2} places an assumption on the actual heterodyne statistics, rather than assuming a trace-distance recovery guarantee. Equation~\eqref{eq:heterodyne-subgaussian} is not implied by trace-distance proximity to a Gaussian state or by finite mean energy alone. It is verified below for the phase-noise models considered here. The parameter $\kappa$ may depend on the input squeezing and displacement; it must not be hidden in a universal constant.

The bound $U$ may be the certified endpoint $U_K$ from Algorithm~\ref{alg:delta-NG}. The numerical certificate is conditioned on the specified physical model; uncertainty in that model must also be included in $U$. If a certificate is itself valid only with failure probability $\delta_{\mathrm{cert}}$, the overall failure probability is at most $\delta+\delta_{\mathrm{cert}}$ for a fixed reconstruction sample size. This is separate from the cost of obtaining the model information or computing the certificate.

The condition $\varepsilon_{\mathrm{tot}}>U$ reserves a positive statistical error budget. In particular, increasing $N$ does not change $\delta_{\mathrm{NG}}(\nu)$. For an exactly Gaussian state one may take $\kappa=1$ and $U=0$, recovering the same dependence on the inverse covariance and accuracy as standard heterodyne tomography, up to universal constants~\cite{bittel2025energy}. A conservative ensemble bound on $\kappa$ need not approach $1$ merely because the phase noise tends to zero.

For our reconstruction procedure, the error bound is the sum of the Gaussian approximation error and the statistical estimation error. For a fixed target state, a consistent moment estimator makes the reconstruction converge to the Gaussian state with the same first moments and covariance matrix as the sample size grows. The remaining error is therefore $\delta_{\mathrm{NG}}(\nu)$. A certified upper bound $U<\varepsilon_{\mathrm{tot}}$ allows us to assign up to $\varepsilon_{\mathrm{tot}}-U$ to the statistical error. If we cannot establish this inequality, the desired accuracy may still be achievable, since $U$ may overestimate the actual approximation error.

For phase-noisy experiments, we must also verify the required tail condition for the actual measurement distribution. Approximate Gaussianity alone does not imply this condition. The following proposition establishes it uniformly over the phase distribution, using bounds on the covariances at fixed phases and on how the corresponding means vary with phase.



\begin{proposition}[Sub-Gaussian statistics under stochastic phase noise]
\label{prop:phase-subgaussian}
Let $\rho_{\mathrm{in}}=\rho_{\mathrm G}(\Sigma_{\mathrm{in}},m_{\mathrm{in}})$, and let each phase realisation act through an orthogonal symplectic matrix $O_\phi$. Set
\begin{equation}
a\coloneqq \lambda_{\min}(\Sigma_{\mathrm{in}})>0,\:
b\coloneqq\lambda_{\max}(\Sigma_{\mathrm{in}}).
\end{equation}
The heterodyne distribution of the ensemble in Eq.~\eqref{Eqphasenoisy} satisfies Eq.~\eqref{eq:heterodyne-subgaussian} with
\begin{equation}
\kappa^2=\frac{1+b+\|m_{\mathrm{in}}\|_2^2}{1+a}.
\label{eq:phase-kappa}
\end{equation}
Moreover, $E\leq2n(1+a^{-1})$, so this quantity is a valid choice of $B$. For a zero-mean input, Eq.~\eqref{eq:phase-kappa} reduces to $\kappa^2=(1+b)/(1+a)$. These statements hold for any probability distribution of the phases, not only in a small-noise approximation.
\end{proposition}

A proof of Theorem~\ref{theorem2}, including the concentration constants, and a proof of Proposition~\ref{prop:phase-subgaussian} are given in Appendix~\ref{Proof of Theorem2}.


\section{Numerical evaluation of the non-Gaussian approximation error}
\label{Numerical evaluation of the non-Gaussian approximation error}
The learning bound requires an upper bound on the error of the moment-matched Gaussian hypothesis.
Here we construct a convergent interval for that error from a specified state or noise model, with explicit control of the discarded Fock-space weight.
This numerical certification is a separate task from estimating moments and the first two moments alone do not supply the required density-matrix information.

\begin{algorithm}[H]
\caption{Adaptive evaluation of $\delta_{\mathrm{NG}}(\nu)$}
\label{alg:delta-NG}
\begin{algorithmic}[1]

\Require State $\varrho_\nu$, moments $(\Sigma,m)$, initial cutoff
$K_0\in\mathbb N_0$, increment $\Delta K\in\mathbb N$, and tolerance $\varepsilon_{\mathrm{tol}}>0$.
\Ensure Certified interval $[L_K,U_K]$ containing
$\delta_{\mathrm{NG}}(\nu)$.

\State Compute the Williamson decomposition
\Statex
$\Sigma
=
S_{\mathrm W}\left(\bigoplus_{j=1}^{n}s_j I_2\right)S_{\mathrm W}^{T}.$

\State Construct the Gaussian state
\Statex
$\rho_{\mathrm G}
=
D(m)U_{S_{\mathrm W}}
\left[
\bigotimes_{j=1}^{n}
\tau\left(\frac{s_j-1}{2}\right)
\right]
U_{S_{\mathrm W}}^\dagger D^\dagger(m).$

\State Set $K\gets K_0$.

\Repeat
    \State Construct the total photon projector $\Pi_K$.

    \State Compute the truncated states
    \Statex 
    $\varrho_{\nu,K}
    =
    \Pi_K\varrho_\nu\Pi_K,
    \:
    \rho_{\mathrm G,K}
    =
    \Pi_K\rho_{\mathrm G}\Pi_K.$

    \State Compute the lower bound
    \Statex
    $L_K
    =
    \frac{1}{2}
    \left\|
    \varrho_{\nu,K}-\rho_{\mathrm G,K}
    \right\|_1.$

    \State Compute the weights
    \Statex 
   $ \eta_{\nu,K}
    =
    1-\operatorname{Tr}(\varrho_{\nu,K}),
    \:
    \eta_{\mathrm G,K}
    =
    1-\operatorname{Tr}(\rho_{\mathrm G,K}).$

    \State Compute the upper bound
    \Statex 
   $ U_K
    =
    \min\left\{
    1,\,
    L_K+\sqrt{\eta_{\nu,K}}+\sqrt{\eta_{\mathrm G,K}}
    \right\}.$

    \If{$U_K-L_K>\varepsilon_{\mathrm{tol}}$}
        \State $K\gets K+\Delta K$.
    \EndIf
\Until{ $U_K-L_K\leq\varepsilon_{\mathrm{tol}}$}

\State \textbf{return} $[L_K,U_K]$.

\end{algorithmic}
\end{algorithm}

Algorithm~\ref{alg:delta-NG} provides a cutoff-based procedure for
evaluating the non-Gaussian approximation error introduced in
Eq.~\eqref{non-gaussian-approximation-error}. Although the moment-matched
Gaussian state $\rho_{\mathrm G}(\Sigma,m)$ is uniquely determined by
$m$ and $\Sigma$, evaluating $\delta_{\mathrm{NG}}(\nu)$ requires
information about the state $\varrho_\nu$. Such information may
be provided by an analytical Fock basis expansion, a stochastic ensemble
from which the density matrix elements can be evaluated, or a numerically
constructed density matrix.

Williamson's decomposition gives a convenient construction of the Gaussian reference density operator.
In the convention where the vacuum covariance matrix is $I_{2n}$, consider
the Williamson decomposition
\begin{equation}
    \Sigma
    =
    S_{\mathrm W}
    \left(
        \bigoplus_{j=1}^{n} s_j I_2
    \right)
    S_{\mathrm W}^{\mathrm T},
    \:
    s_j \geq 1.
\end{equation}
Here, $S_{\mathrm W}$ is a real symplectic matrix satisfying
\begin{equation}
    S_{\mathrm W}\Omega S_{\mathrm W}^{\mathrm T}
    =
    \Omega,
\end{equation}
where $\Omega$ is the symplectic form, and
$s_1,\ldots,s_n$ are the symplectic eigenvalues of $\Sigma$. The
condition $s_j\geq 1$ follows from the uncertainty principle. The
block-diagonal matrix $ \Sigma_{\mathrm{th}}\coloneqq \bigoplus_{j=1}^{n} s_j I_2$
may be interpreted as the covariance matrix of $n$ independent thermal
modes.
For each mode, define the mean thermal occupation number $ \bar n_j= \frac{s_j-1}{2}.$
The corresponding single-mode thermal state is
\begin{equation}
    \tau(\bar n_j)
    =
    \frac{1}{\bar n_j+1}
    \sum_{q=0}^{\infty}
    \left(
        \frac{\bar n_j}{\bar n_j+1}
    \right)^{q}
    \lvert q\rangle\langle q\rvert.
    \label{eq:thermal-state}
\end{equation}
The coefficient multiplying $\lvert q\rangle\langle q\rvert$ is the
probability of finding $q$ photons in the mode. These probabilities sum
to one, and their mean photon number is $\bar n_j$. In the present
covariance convention, $\tau(\bar n_j)$ has zero displacement and
covariance $ (2\bar n_j+1)I_2
    =
    s_j I_2.$
Consequently, the product state $ \rho_{\mathrm{th}}
    \coloneqq
    \bigotimes_{j=1}^{n}\tau(\bar n_j)$
has zero displacement and covariance $\Sigma_{\mathrm{th}}
    =
    \bigoplus_{j=1}^{n}s_j I_2.$

Let $U_{S_{\mathrm W}}$ be the Gaussian unitary implementing the symplectic
transformation $S_{\mathrm W}$, such that $ U_{S_{\mathrm W}}^{\dagger}
    \hat{\boldsymbol R}
    U_{S_{\mathrm W}}
    =
    S_{\mathrm W}\hat{\boldsymbol R},$
where $\hat{\boldsymbol R}$ denotes the vector of quadrature operators.
Finally, let $D(m)$ denote the displacement operator, defined such that
\begin{equation}
    D^{\dagger}(m)
    \hat{\boldsymbol R}
    D(m)
    =
    \hat{\boldsymbol R}+m.
\end{equation}
This displacement changes the first moment vector from zero to $m$
without changing the covariance matrix. The moment-matched Gaussian reference state is therefore
\begin{equation}
        \rho_{\mathrm G}(\Sigma,m)
        =
        D(m)U_{S_{\mathrm W}}
        \left[
            \bigotimes_{j=1}^{n}\tau(\bar n_j)
        \right]
        U_{S_{\mathrm W}}^{\dagger}D^{\dagger}(m).
    \label{eq:rhoGEQ}
\end{equation}
By construction, this state has displacement vector $m$ and covariance
matrix $\Sigma$.
The thermal states appearing in Eq.~\eqref{eq:rhoGEQ} are understood
as canonical mathematical building blocks for the mixed Gaussian reference
state. If $s_j=1$, then $\bar n_j=0$ and $\tau(\bar n_j)
    =
    \lvert 0\rangle\langle 0\rvert;$
hence, the corresponding Williamson mode is simply the vacuum state. If
$s_j>1$, the associated thermal state accounts for the mixedness encoded
by that symplectic eigenvalue.
Alternatively, one may calculate the Gaussian density matrix directly
from the Gaussian characteristic function determined by $(\Sigma,m)$,
or by using direct Fock-space formulas for Gaussian states. In these
equivalent approaches, the thermal states remain implicit rather than
being constructed explicitly~\cite{serafini2023quantum}.

To evaluate the trace norm numerically, we truncate the multimode Fock
space containing at most $K$ photons in total. For
$\boldsymbol q=(q_1,\ldots,q_n)$, define the total
photon number by $ |\boldsymbol q|
    \coloneqq
    \sum_{j=1}^{n}q_j.$
The orthogonal projector onto the truncated subspace is
\begin{equation}
    \Pi_K
    =
    \sum_{|\boldsymbol q|\leq K}
    \lvert\boldsymbol q\rangle\langle\boldsymbol q\rvert.
\end{equation}
Operationally, $\Pi_K$ represents the low-photon-number outcome of the two-outcome measurement $\left\{\Pi_K,\,I-\Pi_K\right\},$
which determines whether the total photon number is at most $K$ or
greater than $K$.

Now, applying the projector to the ensemble state and its moment-matched
Gaussian reference state, we define the operators
\begin{align}
    \varrho_{\nu,K}
    &\coloneqq
    \Pi_K\varrho_\nu\Pi_K,
    &
    \rho_{\mathrm G,K}
    &\coloneqq
    \Pi_K\rho_{\mathrm G}(\Sigma,m)\Pi_K.
\end{align}
These operators retain the matrix elements  within
$|\boldsymbol q|\leq K$, while discarding both the higher photon number
components and the off-diagonal matrix elements connecting the retained
and discarded regions.
They are positive but generally subnormalised, since
\begin{align}
    \operatorname{Tr}(\varrho_{\nu,K})
    &=
    \operatorname{Tr}(\Pi_K\varrho_\nu)
    \leq 1,
    \\
    \operatorname{Tr}(\rho_{\mathrm G,K})
    &=
    \operatorname{Tr}\!\left[
        \Pi_K\rho_{\mathrm G}(\Sigma,m)
    \right]
    \leq 1.
\end{align}
Each trace is the probability that the corresponding state lies within
the retained region; hence, its trace deficit is the probability above
the cutoff. We do not renormalise the compressed operators, because these
deficits are needed to control the truncation error.

Finally, let $ \Delta_K
    \coloneqq
    \varrho_{\nu,K}-\rho_{\mathrm G,K}.$
Since \(\Delta_K\) is a finite-dimensional Hermitian operator, its trace
norm can be evaluated from its eigenvalues \(\{\lambda_\ell(\Delta_K)\}\)
as
\begin{equation}
    \|\Delta_K\|_1
    =
    \sum_{\ell}
    \bigl|\lambda_\ell(\Delta_K)\bigr|.
\end{equation}
This quantity measures the difference between the two compressed operators
within the retained photon number region. The omitted probability weights
can then be combined with the gentle operator lemma to derive an 
upper bound on the remaining truncation error
\cite{winter1999coding,wilde2013sequential}.

Since compressing an operator to a subspace using an orthogonal projector cannot increase its trace norm, the contribution within the retained photon number region satisfies
\begin{align}
2L_K
&\coloneqq
\left\|
\Pi_K
\left[
\varrho_\nu-\rho_{\mathrm G}(\Sigma,m)
\right]
\Pi_K
\right\|_1
\nonumber\\
&\leq
\left\|
\varrho_\nu-\rho_{\mathrm G}(\Sigma,m)
\right\|_1
=
2\delta_{\mathrm{NG}}(\nu).
\end{align}
Consequently, $L_K\leq\delta_{\mathrm{NG}}(\nu),$
so $L_K$ is a lower bound that does not include contributions from
outside the truncated Fock subspace.
The discarded probability weights are
\begin{align}
\eta_{\nu,K}
&\coloneqq
1-\operatorname{Tr}(\varrho_{\nu,K}),
&
\eta_{\mathrm G,K}
&\coloneqq
1-\operatorname{Tr}(\rho_{\mathrm G,K}).
\end{align}
These quantities are the probabilities that the corresponding states
contain more than \(K\) photons in total. Applying the unnormalised form of
the gentle operator lemma
\cite{winter1999coding,wilde2013sequential} gives
\begin{align}
\frac{1}{2}
\left\|
\varrho_\nu-\varrho_{\nu,K}
\right\|_1
&\leq
\sqrt{\eta_{\nu,K}},
\\
\frac{1}{2}
\left\|
\rho_{\mathrm G}(\Sigma,m)-\rho_{\mathrm G,K}
\right\|_1
&\leq
\sqrt{\eta_{\mathrm G,K}}.
\end{align}
The triangle inequality therefore implies
\begin{align}
\delta_{\mathrm{NG}}(\nu)
&\leq
\frac{1}{2}
\left\|
\varrho_\nu-\varrho_{\nu,K}
\right\|_1
+
L_K
+
\frac{1}{2}
\left\|
\rho_{\mathrm G,K}-\rho_{\mathrm G}(\Sigma,m)
\right\|_1
\nonumber\\
&\leq
L_K
+
\sqrt{\eta_{\nu,K}}
+
\sqrt{\eta_{\mathrm G,K}}. 
\end{align}
Since the trace distance between normalised states is at most one, define
\begin{equation}
U_K
\coloneqq
\min\left\{
1,\,
L_K
+
\sqrt{\eta_{\nu,K}}
+
\sqrt{\eta_{\mathrm G,K}}
\right\}.
\end{equation}
Combining the preceding lower and upper estimates yields the certified
interval
\begin{equation}
L_K
\leq
\delta_{\mathrm{NG}}(\nu)
\leq
U_K.
\label{eq:bound-nonG-error}
\end{equation}
Its width satisfies
\begin{equation}
0
\leq
U_K-L_K
\leq
\sqrt{\eta_{\nu,K}}
+
\sqrt{\eta_{\mathrm G,K}}.
\end{equation}
Because the discarded weights vanish as $K\to\infty$, this width
converges to zero. Accordingly, the cutoff is increased until $U_K-L_K
\leq \varepsilon_{\mathrm{tol}},$
where $\varepsilon_{\mathrm{tol}}$ is the prescribed numerical tolerance.

The interval is a model-based certificate, not a test of Gaussianity from the first two moments alone. Its convergence does not establish efficient scaling with the mode number: the retained Fock space has dimension $\binom{n+K}{n}$.

A simple explicit bound follows from the mean total photon number. In our convention,
\begin{equation}
\overline n:=\operatorname{Tr}(\varrho_\nu\widehat N_{\mathrm{ph}})
=\frac{\operatorname{Tr}\Sigma+\|m\|_2^2-2n}{4},
\label{eq:mean-photon-number}
\end{equation}
where $\widehat N_{\mathrm{ph}}=\sum_{j=1}^n a_j^\dagger a_j$ and
$a_j=(R_{2j-1}+iR_{2j})/2$. The Gaussian reference has the same value of $\overline n$. Hence
\begin{equation}
\eta_{\nu,K},\eta_{\mathrm G,K}\leq
\min\{1,\overline n/(K+1)\},
\label{eq:markov-fock-tail}
\end{equation}
and the interval width is at most
$2\sqrt{\min\{1,\overline n/(K+1)\}}$. This follows from the operator inequality
$I-\Pi_K\preceq\widehat N_{\mathrm{ph}}/(K+1)$.

For a numerical implementation, suppose that Hermitian approximations $A_K$ and $B_K$ to $\varrho_{\nu,K}$ and $\rho_{\mathrm G,K}$ satisfy certified trace-norm bounds
\begin{equation}
\|A_K-\varrho_{\nu,K}\|_1\leq\xi_{\nu,K},\qquad
\|B_K-\rho_{\mathrm G,K}\|_1\leq\xi_{\mathrm G,K}.
\end{equation}
Let $\overline\eta_{\nu,K},\overline\eta_{\mathrm G,K}\in[0,1]$ be certified upper bounds on the discarded weights and set
\begin{equation}
\ell_K=\tfrac12\|A_K-B_K\|_1,\qquad
\xi_K=\tfrac12(\xi_{\nu,K}+\xi_{\mathrm G,K}).
\end{equation}
Then the interval with endpoints
\begin{align}
L_K^{\mathrm{num}}&=\max\{0,\ell_K-\xi_K\},\\
U_K^{\mathrm{num}}&=\min\{1,\ell_K+\xi_K+
\sqrt{\overline\eta_{\nu,K}}+
\sqrt{\overline\eta_{\mathrm G,K}}\}
\label{eq:numerical-certificate}
\end{align}
still contains $\delta_{\mathrm{NG}}(\nu)$. The reverse triangle inequality proves
$|\ell_K-L_K|\leq\xi_K$. Any error in the Gaussian reference caused by approximate moments must also be included in $\xi_{\mathrm G,K}$. If the matrix error bounds and tail bounds tend to zero with increasing computational accuracy and cutoff, the numerical interval converges as well. In Algorithm~\ref{alg:delta-NG}, these endpoints replace $L_K,U_K$ whenever the matrix entries are not evaluated exactly.

Substituting the certified estimate $\delta_{\mathrm{NG}}(\nu)\leq U_K$
into
Theorem~\ref{theorem1} gives the model-dependent deterministic reconstruction guarantee
\begin{align}
&D_{\mathrm{tr}}\!\left(
\varrho_\nu,
\widetilde{\rho}_{\mathrm G}
(\widetilde{\Sigma},\widetilde m)
\right)
\nonumber\\
&\leq
\min\Bigg\{
1,\,
U_K
+
\frac{1+\sqrt{3}}{8}
\operatorname{Tr}\!\left[
\left(
\Sigma^{-1}
+
\widetilde{\Sigma}^{-1}
\right)
\left|
\widetilde{\Sigma}-\Sigma
\right|
\right]
\nonumber\\
&\hspace{3.8cm}
+
\frac{1}{2}
\left\|
\widetilde{\Sigma}^{-1/2}
\left(m-\widetilde m\right)
\right\|_2
\Bigg\}.
\label{eq:computable-total-bound}
\end{align}

This expression requires the true moments or valid bounds on their estimation errors and it is not a data-only certificate obtained by substituting empirical moments for unknown quantities. Theorem~\ref{theorem2} supplies the corresponding probabilistic control.

\section{Comparison with existing results}
Our theorems together with Algorithm~\ref{alg:delta-NG} and Theorem~S58 of Mele et al.~\cite{mele2025learning}
address the reconstruction of a state that is approximately Gaussian.
Both use the Gaussian reference with the same first moments and
covariance matrix as the physical state, so that
$G(\varrho_\nu)=\rho_{\mathrm G}(\Sigma,m)$ in our notation. Both also
allow mixed states and use the same error decomposition,
\begin{equation}
    D_{\mathrm{tr}}(\varrho_\nu,\widetilde\rho_{\mathrm G})
    \leq
    \delta_{\mathrm{NG}}(\nu)
    +
    D_{\mathrm{tr}}\!\left(
        \rho_{\mathrm G}(\Sigma,m),\widetilde\rho_{\mathrm G}
    \right).
\end{equation}
Here, $\delta_{\mathrm{NG}}(\nu)$ is the error incurred by replacing
the output $\varrho_\nu$ with its moment-matched Gaussian reference state $\rho_{\mathrm G}(\Sigma,m)$. The second term is the
statistical reconstruction error caused by estimating the first
moments and covariance matrix from finite measurement data.
The differences concern how each of these two terms is controlled.
Both analyses give guarantees for reconstructing a Gaussian state from estimates of the first moments and covariance matrix. We therefore compare the information they use to bound the Gaussian approximation error and the assumptions about measurement outcomes that determine how the statistical error decreases as the sample size grows.

The first difference concerns the Gaussian approximation error.
For a target accuracy $\varepsilon_{\mathrm{tot}}$, Theorem~S58 assumes
that the relative entropy of non-Gaussianity satisfies
\begin{equation}
    d_{\mathcal G}(\varrho_\nu)
    =S\!\left(\varrho_\nu\|
        \rho_{\mathrm G}(\Sigma,m)\right)
    \leq\varepsilon_{\mathrm{tot}}^2,
    \label{eq:relative}
\end{equation}
where $S(\rho\|\sigma)$ is the quantum relative entropy with logarithms to base two. Pinsker's inequality then gives
\begin{equation}
    \delta_{\mathrm{NG}}(\nu)
    \leq
    \sqrt{\frac{\ln2}{2}\,
        d_{\mathcal G}(\varrho_\nu)}
    \leq
    \sqrt{\frac{\ln2}{2}}\,
        \varepsilon_{\mathrm{tot}}.
    \label{eq:comparison-pinsker}
\end{equation}
Thus, Mele et al. obtain an upper bound on the approximation error from
an assumed relative entropy bound and reserve the remaining fraction
of $\varepsilon_{\mathrm{tot}}$ for statistical reconstruction.

In our approach, Algorithm~1 evaluates the
trace-distance approximation error for a specified output state or
phase-noise model. It constructs the density operators within a total
photon number cutoff and bounds the omitted contributions, giving
\begin{equation}
    L_K\leq\delta_{\mathrm{NG}}(\nu)\leq U_K,
    \: U_K-L_K\longrightarrow0
    \:\text{as }K\to\infty.
\end{equation}

This provides lower and upper bounds on the Gaussian
approximation error, with numerical precision controlled by the photon
number cutoff. It complements the relative-entropy approach of
Mele et al. by providing a numerical evaluation for a specified noise
model. If $U_K<\varepsilon_{\mathrm{tot}}$, we can allocate at most
$U_K$ to the Gaussian approximation error and require the statistical
reconstruction error to be at most
$\varepsilon_{\mathrm{tot}}-U_K$.

Increasing $K$ reduces the uncertainty caused by omitting higher photon
number components. As their probability weights vanish, the interval
$[L_K,U_K]$ narrows around the actual value of
$\delta_{\mathrm{NG}}(\nu)$. The cutoff calculation and Pinsker's
inequality provide complementary upper bounds: for a given state and
cutoff, either bound may be smaller. When both are available, the
smaller bound can be used in the reconstruction guarantee.

The interval above accounts for the error caused by truncating the
Fock space. If the density-matrix elements within the truncated
subspace are also computed approximately, for example through
numerical integration over the phase distribution, those numerical
errors must be bounded and included in the interval to preserve
its certification.

The second difference concerns the statistical reconstruction error.
The Gaussian estimate underlying Theorem~S58 has a square-root
dependence on covariance error. Theorem~1 instead uses
the later Gaussian perturbation bounds of
Refs.~\cite{bittel2025optimal,bittel2025energy}, which are linear in
the moment errors when the covariance inverses remain bounded.
Consequently, moment errors of order $N^{-1/2}$ lead to state error
bounds of order $N^{-1/4}$ in the earlier analysis and $N^{-1/2}$ in
ours, with the corresponding state-dependent factors. 

More explicitly, under its stated relative-entropy and energy assumptions, Theorem~S58 states
\begin{equation}
    N_{\mathrm{S}}
    =\mathcal O\!\left(
        \frac{n^7E_{\mathrm{av}}^4}
        {\varepsilon_{\mathrm{tot}}^4}
        \log\!\frac{n^2}{\delta}
        \right),
\end{equation}
where $E_{\mathrm{av}}$ bounds the mean energy per mode. Under the sub-Gaussian assumption of Theorem~\ref{theorem2}, and with a known bound $B\geq2n+\operatorname{Tr}(\Sigma^{-1})$, our certified approximation bound gives the sufficient sample scaling
\begin{equation}
N=\mathcal O\!\left(
\frac{\kappa^4[n+\ln(4/\delta)]B^2}
{(\varepsilon_{\mathrm{tot}}-U_K)^2}
\right),\: U_K<\varepsilon_{\mathrm{tot}}.
\label{eq:comparison-certified-samples}
\end{equation}

The covariance factor $2n+\operatorname{Tr}(\Sigma^{-1})$, denoted
by $E$ in Theorem~2, has a different meaning from
$E_{\mathrm{av}}$. If a known lower bound on the smallest covariance eigenvalue is independent of $n$, one may choose $B=\mathcal O(n)$. If also $\kappa=\mathcal O(1)$, this gives an $\mathcal O\!\left(n^3/(\varepsilon_{\mathrm{tot}}-U_K)^2\right)$ sample bound at fixed failure probability. The two sample bounds use different promises and measurement analyses: this comparison concerns their accuracy exponents and is not an unconditional improvement under identical hypotheses.


\subsection{Relation to doped Gaussian-state learning}
\label{sec:doped-learning}

A complementary way to go beyond Gaussian tomography is to enlarge the hypothesis class using structure in the preparation circuit.
In the fermionic setting, Mele and Herasymenko learn states prepared with arbitrary Gaussian gates and at most $t$ non-Gaussian gates, each supported on a bounded number of Majorana operators, with sample and time costs polynomial in $n$ and $2^t$ at fixed accuracy~\cite{meleherasymenko2025doped}.
For fixed gate locality this permits $t=O(\log n)$.
A direct bosonic analogue is established in Ref.~\cite{mele2025learning}: for pure states prepared from the vacuum with $t$ non-Gaussian gates, each generated by a polynomial in at most $\ell$ quadrature operators, Gaussian processing compresses the non-Gaussian part onto at most $\ell t$ modes.
We use $\ell$ for gate locality to distinguish it from the heterodyne tail parameter $\kappa$ of Theorem~\ref{theorem2}.
Under the second energy moment promise in that work and at fixed confidence, a sufficient sample bound has the form $\operatorname{poly}(n,E_{\mathrm{dop}},\epsilon^{-1})+(nE_{\mathrm{dop}}/\epsilon)^{O(\ell t)}$, where $E_{\mathrm{dop}}$ bounds the square root of the second energy moment per mode.
Thus constant $\ell t$ gives polynomial learning costs; the bosonic bound does not extend the fermionic logarithmic-gate regime because each active bosonic mode is infinite-dimensional.

For pure states, symplectic rank characterises the minimum number of non-Gaussian active modes under Gaussian compression~\cite{mele2026symplecticrank}.
For mixed states, such a compression is an additional structural promise.
Suppose, more generally, that a state with finite second moments has the factorisation
\begin{equation}
\rho=U_{\mathrm G}(\tau_k\otimes\omega_{\mathrm G})U_{\mathrm G}^{\dagger},
\label{eq:mixed-compression-promise}
\end{equation}
where $U_{\mathrm G}$ is Gaussian, $\tau_k$ is an arbitrary $k$-mode state, and $\omega_{\mathrm G}$ is Gaussian on the remaining modes.
Writing $G(\rho)$ for the moment-matched Gaussian reference, covariance of moment matching under Gaussian unitaries and its compatibility with tensor products give
\begin{equation}
G(\rho)=U_{\mathrm G}\bigl(G(\tau_k)\otimes\omega_{\mathrm G}\bigr)U_{\mathrm G}^{\dagger}.
\label{eq:compressed-gaussian-reference}
\end{equation}
Unitary invariance and multiplicativity of the trace norm under tensoring with a state then imply the exact identity
\begin{equation}
D_{\mathrm{tr}}\bigl(\rho,G(\rho)\bigr)
=D_{\mathrm{tr}}\bigl(\tau_k,G(\tau_k)\bigr).
\label{eq:compressed-approximation-error}
\end{equation}
This elementary observation localises the intrinsic approximation error entirely to the active modes.
If the Gaussian frame and the reduced state are available from the model, Algorithm~\ref{alg:delta-NG} can therefore be applied on $k$ modes, with truncated dimension $\binom{k+K}{k}$ rather than $\binom{n+K}{n}$.
The photon-number tails must be bounded in that frame, since an active Gaussian transformation need not preserve energy.
Learning an unknown frame or the non-Gaussian reduced state entails additional costs; the identity alone is not a tomography algorithm.

The phase-noisy and accepted UA ensembles illustrate a different source of non-Gaussianity: they mix Gaussian conditional states.
Such mixing can give $\delta_{\mathrm{NG}}>0$ without introducing genuine quantum non-Gaussianity outside the convex Gaussian hull.
Indeed, the mixed-state extension of symplectic rank vanishes on mixtures of Gaussian states~\cite{mele2026symplecticrank}.
This does not imply a common low-mode factorisation of the mixed state as in Eq.~\eqref{eq:mixed-compression-promise}.
Consequently, a nonzero approximation error is not a measure of the number of non-Gaussian gates, and a few-gate learning theorem cannot be inferred from approximate Gaussianity alone.
Under an independently justified compression promise, however, learning the active non-Gaussian state would remove the corresponding Gaussian-model error floor, while Eq.~\eqref{eq:compressed-approximation-error} already reduces the cost of certifying that floor.

\section{Example: Trace-distance bound for a unitary-averaged state}

We illustrate Theorem~\ref{theorem1} using the unitary-averaging (UA)
protocol of Ref.~\cite{swain2024improving}, reviewed in
Appendix~\ref{Unitary Averaging in a nutshell}. In this protocol, one arm
of a two-mode squeezed vacuum is distributed across
$n_{\mathrm{UA}}$ independently phase-noisy channel uses. After decoding,
the $(n_{\mathrm{UA}}-1)$ ancillary output modes are projected onto the
vacuum, leaving a heralded two-mode output state.

Let $\boldsymbol{\phi}
=
(\phi_1,\ldots,\phi_{n_{\mathrm{UA}}})$
denote a realisation of the independent phase channels. Following
Ref.~\cite{swain2024improving}, define
\begin{align}
\alpha(\boldsymbol{\phi})
e^{i\phi_\beta(\boldsymbol{\phi})}
&\coloneqq
\frac{1}{n_{\mathrm{UA}}}
\sum_{k=1}^{n_{\mathrm{UA}}}e^{i\phi_k},\\
x(\boldsymbol{\phi})
&\coloneqq
\tanh r'(\boldsymbol{\phi})
=
\alpha(\boldsymbol{\phi})\tanh r.
\end{align}
We take $r\geq0$ and $n_{\mathrm{UA}}\in\mathbb N$. Then $0\leq\alpha\leq1$, so $0\leq r'\leq r$. If $\alpha=0$, the phase $\phi_\beta$ can be chosen arbitrarily because $x=0$. For a fixed phase realisation, the normalised heralded output is the two-mode squeezed state
\begin{equation}
\lvert\psi_{\mathrm{out}}(\boldsymbol{\phi})\rangle
=
\sqrt{1-x(\boldsymbol{\phi})^2}
\sum_{\ell=0}^{\infty}
\left[
-x(\boldsymbol{\phi})
e^{i\phi_\beta(\boldsymbol{\phi})}
\right]^\ell
\lvert\ell,\ell\rangle.
\label{eq:UA-fixed-output}
\end{equation}
The corresponding success probability is
\begin{equation}
P_{\mathrm{succ}}(\boldsymbol{\phi})
=
\left[
\frac{\cosh r'(\boldsymbol{\phi})}{\cosh r}
\right]^2. \label{eq:SuccProbUA}
\end{equation}

Let $p_{\nu,n_{\mathrm{UA}}}(\boldsymbol\phi)$ be the phase density before heralding. The average success probability and the accepted phase density are
\begin{align}
\overline P_{\mathrm{succ}}&:=\int p_{\nu,n_{\mathrm{UA}}}(\boldsymbol\phi)
P_{\mathrm{succ}}(\boldsymbol\phi)\,d\boldsymbol\phi,\label{eq:average-UA-success}\\
p_{\nu,n_{\mathrm{UA}}}^{\mathrm{acc}}(\boldsymbol\phi)
&=\frac{p_{\nu,n_{\mathrm{UA}}}(\boldsymbol\phi)P_{\mathrm{succ}}(\boldsymbol\phi)}
{\overline P_{\mathrm{succ}}}.
\label{eq:accepted-phase-law}
\end{align}
The same definitions apply to general probability measures. Since $0\leq r'\leq r$, we have
$\operatorname{sech}^2r\leq P_{\mathrm{succ}}(\boldsymbol\phi)\leq1$, so the conditional law is well defined for finite $r$.

Let
$p_{\nu,n_{\mathrm{UA}}}^{\mathrm{acc}}(\boldsymbol{\phi})$ denote the
phase distribution conditioned on successful vacuum heralding. The
 output state is then
\begin{equation}
\varrho_\nu^{\mathrm{UA}}
=
\int
p_{\nu,n_{\mathrm{UA}}}^{\mathrm{acc}}(\boldsymbol{\phi})\,
\lvert\psi_{\mathrm{out}}(\boldsymbol{\phi})\rangle
\langle\psi_{\mathrm{out}}(\boldsymbol{\phi})\rvert\,
d\boldsymbol{\phi}.
\label{eq:UA-ensemble-output}
\end{equation}
Because both $x(\boldsymbol{\phi})$ and
$\phi_\beta(\boldsymbol{\phi})$ fluctuate between experimental runs,
$\varrho_\nu^{\mathrm{UA}}$ is generally non-Gaussian.

The first moments vanish for every phase distribution. The phase distribution before heralding describes
how the channel phases $\boldsymbol{\phi}$ are distributed across all
extra modes. The conditional phase distribution in
Eq.~\eqref{eq:accepted-phase-law} describes those same phases among the
attempts in which all ancillary output modes are found in the vacuum. To simplify the covariance matrix, assume that the phase distribution
before heralding is invariant under
$\boldsymbol{\phi}\mapsto-\boldsymbol{\phi}$. Since the probability of
successful heralding is the same for $\boldsymbol{\phi}$ and
$-\boldsymbol{\phi}$, conditioning on success preserves this symmetry.
The sine-dependent correlations change sign under this transformation
and therefore average to zero. Define
\begin{align}
D
&\coloneqq
\mathbb{E}\!\left[
\frac{1+x(\boldsymbol{\phi})^2}
     {1-x(\boldsymbol{\phi})^2}
\right],
\\
C
&\coloneqq
\mathbb{E}\!\left[
\frac{2x(\boldsymbol{\phi})}
     {1-x(\boldsymbol{\phi})^2}
\cos\!\bigl(\phi_\beta(\boldsymbol{\phi})\bigr)
\right],
\label{eq:UA-C-exact}
\end{align}
where the expectation is taken over the accepted runs. In the quadrature
ordering $(q_1,p_1,q_2,p_2)$, the covariance matrix of
$\varrho_\nu^{\mathrm{UA}}$ is
\begin{equation}
\Sigma
=
\begin{pmatrix}
D & 0 & -C & 0 \\
0 & D & 0 & C \\
-C & 0 & D & 0 \\
0 & C & 0 & D
\end{pmatrix}.
\label{eq:UA-true-covariance}
\end{equation}
The signs of the correlation block follow the convention of
Ref.~\cite{swain2024improving}; a local phase convention can reverse both
signs without affecting the bound below, which depends only on $C^2$.

No small-noise approximation is required for the results below. All covariance entries, matrix elements, and tail probabilities are evaluated using the exact conditional ensemble in Eq.~\eqref{eq:UA-ensemble-output}. In particular, phase averages are taken after weighting by the heralding probability. Replacing these averages by functions of mean parameters would require separate remainder bounds.
The Fock basis form in Eq.~\eqref{eq:UA-fixed-output} can be used directly
to evaluate the non-Gaussian approximation error. For a total photon
cutoff $K$, let
\begin{equation}
M_K
\coloneqq
\left\lfloor\frac{K}{2}\right\rfloor.
\end{equation}
The truncated output is
\begin{align}
\varrho_{\nu,K}^{\mathrm{UA}}
=
\sum_{\ell,\ell'=0}^{M_K}
(-1)^{\ell+\ell'}
\mathbb{E}\!\Big[
&
\bigl(1-x^2\bigr)
x^{\ell+\ell'}
e^{i(\ell-\ell')\phi_\beta}
\Big]
\nonumber\\[-1mm]
&\times
\lvert\ell,\ell\rangle
\langle\ell',\ell'\rvert,\label{eq:UA-truncated-density}
\end{align}
where the dependence of $x$ and $\phi_\beta$ on
$\boldsymbol{\phi}$ is implicit. Its discarded probability weight is
\begin{equation}
\eta_{\nu,K}^{\mathrm{UA}}
=
\mathbb{E}\!\left[
x^{2(M_K+1)}
\right].
\label{eq:UA-discarded-weight}
\end{equation}
Physically, $\eta_{\nu,K}^{\mathrm{UA}}$ is the probability that an
accepted UA output lies outside the truncated Fock subspace. Since
$\lvert\ell,\ell\rangle$ contains $2\ell$ photons in total, the cutoff
keeps only $\ell\leq M_K$. For each accepted phase realisation, the
discarded geometric tail is $x^{2(M_K+1)}$, and the expectation accounts
for the stochastic variation of the squeezing between accepted
runs.

The Gaussian reference must be compressed on the full total photon number subspace, not only on the span of the vectors $|\ell,\ell\rangle$. Although the physical output has this pair-number support, its Gaussian reference generally does not.

For any distribution of the channel phases, the tail satisfies
\begin{equation}
\eta_{\nu,K}^{\mathrm{UA}}\leq(\tanh r)^{2(M_K+1)}.
\end{equation}
The mean total photon number of the output and of its Gaussian reference is at most $2\sinh^2r$. Equation~\eqref{eq:markov-fock-tail} therefore gives
\begin{align}
U_K^{\mathrm{UA}}-L_K^{\mathrm{UA}}
\leq{}&(\tanh r)^{M_K+1}\nonumber\\
&+\sqrt{\min\{1,2\sinh^2r/(K+1)\}}.
\label{eq:UA-explicit-cutoff-width}
\end{align}
This supplies an explicit convergence bound in addition to the exact tail expression.

Combining Eq.~\eqref{eq:UA-truncated-density} with the truncated
 Gaussian state with the same first moments and covariance matrix gives the certified upper endpoint $\delta_{\mathrm{NG}}(\nu)
\leq
U_K^{\mathrm{UA}},$
according to the procedure of
Section~\ref{Numerical evaluation of the non-Gaussian approximation error}. Here
$U_K^{\mathrm{UA}}$ denotes the quantity $U_K$ evaluated for the
unitary-averaged output state. 

As a deterministic illustration of Theorem~\ref{theorem1}, consider an error affecting only the local variances of the transmitted mode, with estimated covariance
\begin{equation}
\widetilde{\Sigma}
=
\begin{pmatrix}
D & 0 & -C & 0 \\
0 & D & 0 & C \\
-C & 0 & D+\epsilon_\Sigma & 0 \\
0 & C & 0 & D+\epsilon_\Sigma
\end{pmatrix},
\:
\epsilon_\Sigma>0,
\label{eq:UA-estimated-covariance}
\end{equation}
and take $m=\widetilde m=0$. Hence, the displacement contribution in
Theorem~\ref{theorem1} vanishes. The covariance perturbation is
\begin{equation}
\Delta\Sigma
\coloneqq
\widetilde{\Sigma}-\Sigma
=
\epsilon_\Sigma
\operatorname{diag}(0,0,1,1).
\end{equation}
Since $\Delta\Sigma\succeq0$, $\left|
\widetilde{\Sigma}-\Sigma
\right|
=
\Delta\Sigma.$
The relevant diagonal elements of the inverse covariance matrices are
\begin{align}
\left(\Sigma^{-1}\right)_{3,3}
=
\left(\Sigma^{-1}\right)_{4,4}
&=
\frac{D}{D^2-C^2},
\\
\left(\widetilde{\Sigma}^{-1}\right)_{3,3}
=
\left(\widetilde{\Sigma}^{-1}\right)_{4,4}
&=
\frac{D}{D^2+D\epsilon_\Sigma-C^2}.
\end{align}
It follows that
\begin{align}
&\operatorname{Tr}\!\left[
\left(
\Sigma^{-1}
+
\widetilde{\Sigma}^{-1}
\right)
\left|
\widetilde{\Sigma}-\Sigma
\right|
\right] \nonumber\\
&=
2\epsilon_\Sigma D
\left(
\frac{1}{D^2-C^2}
+
\frac{1}{D^2+D\epsilon_\Sigma-C^2}
\right).
\end{align}

Substituting this expression and
$\delta_{\mathrm{NG}}(\nu)\leq U_K^{\mathrm{UA}}$ into
Theorem~\ref{theorem1} gives the certified trace-distance bound
\begin{align}
&D_{\mathrm{tr}}\!\left(
\varrho_\nu^{\mathrm{UA}},
\widetilde{\rho}_{\mathrm G}
\right) \nonumber\\
&\leq
\min\Bigg\{
1,\,
U_K^{\mathrm{UA}} \nonumber\\
&
+
\frac{1+\sqrt{3}}{4}\,
\epsilon_\Sigma D
\left(
\frac{1}{D^2-C^2}
+
\frac{1}{D^2+D\epsilon_\Sigma-C^2}
\right)
\Bigg\}.
\label{singlemodeUATrace}
\end{align}

Equation~\eqref{singlemodeUATrace} separates the reconstruction error into
a certified upper bound on the intrinsic non-Gaussian contribution and a
covariance estimation contribution. The averaging order
$n_{\mathrm{UA}}$ affects both terms: it changes the distribution of the
effective squeezing and phase entering $U_K^{\mathrm{UA}}$, while also
modifying the averaged covariance parameters $D$ and $C$. The restricted perturbation in Eq.~\eqref{eq:UA-estimated-covariance} is an illustration, not the form of a general empirical covariance error. The statistical result below uses the full physical estimator from Theorem~\ref{theorem2}, whose errors need not be confined to these entries.

The sample size in Theorem~\ref{theorem2} counts heterodyne outcomes from independent, successfully heralded copies. The following consequence verifies its assumptions for the UA ensemble, without a Gaussian approximation to the measurement statistics.

\begin{corollary}[Certified tomography of the unitary-averaged output.]
\label{cor:UA-tomography}
Consider the state in Eq.~\eqref{eq:UA-ensemble-output}, with fixed $r\geq0$ and independent repetitions of the protocol. Its first moments vanish, and its heterodyne distribution satisfies Eq.~\eqref{eq:heterodyne-subgaussian} with
\begin{equation}
\kappa^2=e^{2r}.
\label{eq:UA-subgaussian-parameter}
\end{equation}
Let $\Sigma$ be its exact covariance, and choose a known bound
\begin{equation}
B\geq4+\operatorname{Tr}(\Sigma^{-1}).
\end{equation}
A choice valid for every phase distribution and every averaging order is
\begin{equation}
B=4(1+e^{2r}).
\label{eq:UA-explicit-B}
\end{equation}
If $U_K^{\mathrm{UA}}$ is a certified upper bound for the moment-matched Gaussian approximation error and $U_K^{\mathrm{UA}}<\varepsilon_{\mathrm{tot}}\leq1$, collecting
\begin{equation}
N_{\mathrm{acc}}=
\left\lceil
C_*\frac{e^{4r}[4\ln9+\ln(4/\delta)]B^2}
{(\varepsilon_{\mathrm{tot}}-U_K^{\mathrm{UA}})^2}
\right\rceil
\label{eq:UA-accepted-sample-complexity}
\end{equation}
accepted outcomes is sufficient for trace-distance error at most $\varepsilon_{\mathrm{tot}}$ with probability at least $1-\delta$. Here $C_*$ is the universal constant from Theorem~\ref{theorem2}, and the reconstruction is the estimator specified there.
\end{corollary}

\noindent\textit{Proof.}
For fixed phases, Eq.~\eqref{eq:UA-fixed-output} is a zero-mean Gaussian state whose covariance eigenvalues are $e^{2r'},e^{2r'},e^{-2r'},e^{-2r'}$. Since $0\leq\alpha\leq1$, we have $0\leq r'\leq r$. Writing $H_\phi=\Sigma_\phi+I_4$, this gives
\begin{equation}
(1+e^{-2r})I_4\preceq H_\phi\preceq(1+e^{2r})I_4.
\end{equation}
The covariance of the accepted heterodyne ensemble is $H=\mathbb E_{\mathrm{acc}}[H_\phi]$, so
$H_\phi\preceq e^{2r}H$. Consequently,
\begin{equation}
\mathbb E[e^{u^{\mathrm T}X}]
=\mathbb E_{\mathrm{acc}}[e^{u^{\mathrm T}H_\phi u/2}]
\leq e^{e^{2r}u^{\mathrm T}Hu/2}.
\end{equation}
Also, $\Sigma=\mathbb E_{\mathrm{acc}}[\Sigma_\phi]\succeq e^{-2r}I_4$, which proves Eq.~\eqref{eq:UA-explicit-B}. Theorem~\ref{theorem2}, with $n=2$, proves the claim. \hfill$\square$

If independent trials are repeated until the fixed number $N_{\mathrm{acc}}$ in Eq.~\eqref{eq:UA-accepted-sample-complexity} of successes has been obtained, the accepted states are independent copies of $\varrho_\nu^{\mathrm{UA}}$ and
\begin{equation}
\mathbb E[N_{\mathrm{att}}]
=\frac{N_{\mathrm{acc}}}{\overline P_{\mathrm{succ}}}.
\label{eq:UA-attempt-sample-complexity}
\end{equation}
Indeed, the waiting times between successive successes are independent geometric random variables with mean $1/\overline P_{\mathrm{succ}}$, and each accepted output has the conditional phase law defined above. Thus the sufficient expected cost is
\begin{equation}
\mathbb E[N_{\mathrm{att}}]=\mathcal O\!\left(
\frac{e^{4r}[1+\ln(4/\delta)]B^2}
{\overline P_{\mathrm{succ}}
(\varepsilon_{\mathrm{tot}}-U_K^{\mathrm{UA}})^2}\right).
\label{eq:UA-expected-cost}
\end{equation}
Each attempt uses $n_{\mathrm{UA}}$ noisy channel calls; their expected number is therefore
$n_{\mathrm{UA}}N_{\mathrm{acc}}/\overline P_{\mathrm{succ}}$.


\section{Discussion and conclusion}

In this work, we have presented a trace-distance reconstruction bound for approximately
Gaussian mixed states produced by stochastic phase noise channels. Although each phase realisation preserves Gaussianity, averaging over the phase fluctuations generally
produces a non-Gaussian mixed output state.  As in the robustness analysis of Ref.~\cite{mele2025learning},
the total reconstruction error is bounded by the sum of the
Gaussian approximation error and the statistical reconstruction
error arising from estimation of the first moments and covariance
matrix.

Theorem~\ref{theorem1} uses recent Gaussian trace-distance
bounds~\cite{bittel2025optimal,bittel2025energy} to control covariance
and displacement estimation errors. When the covariance inverses
remain bounded, this contribution is linear in the moment errors,
improving the square-root dependence on covariance error in the
estimate of Ref.~\cite{mele2025learning}.
Our framework combines these Gaussian trace-distance bounds with a certified
evaluation of the approximation error for a specific stochastic phase noise model,
complementing the relative entropy approach to robustness.

The Fock-space truncation procedure provides certified lower and upper bounds on the intrinsic Gaussian approximation error that converge to its actual value as the cutoff increases. When the upper bound is smaller than the target total error, the difference determines the maximum allowed statistical reconstruction error. Under the explicit sub-Gaussian assumption, the sufficient number of heterodyne samples scales inversely with the square of this allowed error, with the mode number, covariance bound, tail parameter, and confidence level fixed. Proposition~\ref{prop:phase-subgaussian} and Corollary~\ref{cor:UA-tomography} verify that assumption for the phase-noise and accepted UA ensembles. This connects the residual phase noise to the number of measurements needed to achieve the desired accuracy.

Our result provides a learning guarantee within the Gaussian hypothesis class, meaning that the estimator is restricted to returning Gaussian states. Model misspecification arises when phase noise makes the actual target state non-Gaussian, so that it cannot be represented exactly within this family. More data reduce the statistical estimation error, but the reconstruction still approaches the Gaussian state with the target’s first moments and covariance matrix. The remaining approximation error can be reduced by changing the state preparation or the reconstruction rule, for example by allowing non-Gaussian outputs. The model-based approximation certificate bounds the distance between the actual target and its Gaussian reference, while the concentration guarantee bounds the uncertainty in the moments estimated from measurement data. 
Extending the estimator to include selected non-Gaussian features would raise a natural model-selection question: how much additional sample cost is needed to reduce the approximation error?

Finally, we applied the Fock-space truncation to bound the Gaussian approximation error of the heralded two-mode output from the unitary averaging protocol in Ref.~\cite{swain2024improving}. We also evaluated how errors in estimating the covariance matrix affect the reconstruction accuracy.
The UA success probability does not alter the conditional state trace
bound, but it increases the expected number of protocol attempts required
to obtain the necessary accepted heterodyne samples.

More efficient Fock-space calculations and sharper concentration constants would help extend the framework to multimode systems and a wider range of noise channels~\cite{swain2025noise}. The bounds established here are sufficient guarantees and we do not claim optimal sample complexity.

\section{Acknowledgements}
We gratefully acknowledge  Francesco Anna Mele for his constructive feedback, and valuable contributions towards improving the mathematical rigour and clarity of the manuscript. In particular, Francesco used his access to ChatGPT-6 Astra Pro to assist in identifying and correcting mathematical inaccuracies, formulating revised mathematical statements, and developing their corresponding derivations. The authors take full responsibility for the final content of the manuscript. JE and NW acknowledge support from the BMFTR (PhoQuant and QPIC-1). JE has been supported by the BMFTR (PasQuops,QSolid), Berlin Quantum, the Munich Quantum Valley, theClusters of Excellence ML4Q and MATH+, the DFG (CRC183 and SPP 2541), and the European Research Council (ERCAdG DebuQC).
SNS was supported by the Sydney Quantum Academy, Sydney, NSW, Australia.

\begin{widetext}

\appendix
\section{Gaussian noise channels and conventions} \label{Gaussian Noise channel}
We collect the phase-space conventions used in the reconstruction bounds and their proofs.
In particular, we fix the vacuum covariance and displacement normalisation, recall physicality of covariance matrices, and describe the additive Gaussian noise channel used to regularise boundary covariances.

Let $\hat{x}_{1}, \hat{p}_{1}, \dots, \hat{x}_{n}, \hat{p}_{n}$ denote the standard position and momentum operators associated with each mode \cite{serafini2023quantum}. The vector of quadrature operators is then defined as

\begin{align}
   \hat{R} = (\hat{x}_{1}, \hat{p}_{1},\dots, \hat{x}_{n}, \hat{p}_{n})^{T}, 
\end{align}
and it satisfies the 
\emph{canonical commutation relations} (CCR), captured in the symplectic form 
\begin{align}
    \Omega = \bigoplus_{i =1}^{n}\begin{pmatrix}
        0 & 1 \\
        -1 &0
    \end{pmatrix}.
\end{align}
The characteristic function $X_{\alpha}: \mathbb{R}^{2n} \to \mathbb{C}$ associated with a trace-class operator $\alpha$ on $L^{2}(\mathbb{R}^{n})$ is defined as
\begin{align}
    X_{\alpha}(r) =\Tr[\alpha \hat{D}_{r}],
\end{align}

where the displacement operator is defined as 
\begin{align}
    \hat{D}_{r} = e^{-i r^{T} \Omega \hat{R}/2}.
\end{align}

A trace-class operator $\alpha$ has the Weyl expansion
\begin{align}
    \alpha = W(X_{\alpha}(r)) = \int_{\mathbb{R}^{2n}} \frac{\,dr}{(4 \pi)^{n}} X_{\alpha}(r) \hat{D}_{r}^{\dag}. 
\end{align}

This integral is understood as a Hilbert--Schmidt limit over expanding bounded domains, rather than as an absolutely convergent integral for every trace-class operator. The Weyl transform uniquely determines the operator.
 The first moment $m(\rho)$ and covariance matrix $\Sigma(\rho)$ of a quantum state $\rho$ are defined by
\begin{align}
   m(\rho) & = \Tr[\hat{R}\rho]  \\
   \Sigma(\rho) & = \frac{1}{2} \Tr[\{(\hat{R} -m(\rho)\hat{\mathbb{I}}), (\hat{R} -m(\rho)\hat{\mathbb{I}})^{T}\}\rho], 
\end{align}
where $(\cdot)^\top$ denotes the transpose operation, and 
$\{\hat{A}, \hat{B}\} \coloneqq \hat{A}\hat{B} + \hat{B}\hat{A}$ denotes the anticommutator.

Any covariance matrix $\Sigma$ admits the Williamson decomposition~\cite{weedbrook2012gaussian, serafini2023quantum}
\begin{align}
   \Sigma = S D S^{\top}, \label{Williamson}
\end{align}

where $S$ is a symplectic matrix, i.e., 
\begin{align}
    S \Omega S^{\top} = \Omega,
\end{align}
and $D$ is a diagonal matrix of the form
\begin{align}
    D = \bigoplus_{j=1}^{n}
\begin{pmatrix}
d_j & 0 \\
0 & d_j
\end{pmatrix},
\end{align}
with $d_1, d_2, \ldots, d_n \ge 1$.

Any covariance matrix $\Sigma(\rho)$ satisfies the uncertainty relation
\begin{align}
    \Sigma(\rho) + i\Omega \ge 0.
\end{align}

Every finite physical covariance matrix is in fact positive definite. Indeed, if a real vector $v$ had $v^{\mathrm T}\Sigma v=0$, positivity of $\Sigma+i\Omega$ would imply $(\Sigma+i\Omega)v=0$, and hence $\Omega v=0$; the symplectic form is invertible. Conversely, every real symmetric $2n\times2n$ matrix $\Gamma$ satisfying
\begin{align}
    \Gamma + i\Omega \ge 0 
\end{align}

is the covariance matrix of an $n$-mode Gaussian state. A displacement transformation shifts the first moments without affecting the covariance matrix, i.e.,

\begin{align}
   m\!\left(\hat{D}_{\mathbf r}\,\rho\,\hat{D}_{\mathbf r}^\dagger\right)
= m(\rho) + \mathbf r,
\quad
\Sigma\!\left(\hat{D}_{\mathbf r}\,\rho\,\hat{D}_{\mathbf r}^\dagger\right)
= \Sigma(\rho), 
\end{align}
because
\begin{align}
    \hat{D}^\dagger_{\mathbf r}\,\hat{\mathbf R}\,\hat{D}_{\mathbf r}
= \hat{\mathbf R} + \mathbf r\,\hat{\mathbb I}.
\end{align}

Let $M$ be a $2n \times 2n$ positive semidefinite matrix. The Gaussian noise channel $\mathcal{N}_{M}$ is given by a convex combination of displacement transformations $\rho\mapsto\hat D_u\rho\hat D_u^\dagger$, where $u \in \mathbb{R}^{2n}$ follows a Gaussian probability distribution with zero mean and covariance matrix $M$ \cite{serafini2023quantum}. Explicitly, $\mathcal{N}_{M}$ can be written as
\begin{align}
    \mathcal{N}_{M}(\varphi )  = \int \,du  \mathcal{P}_{M}(u) \hat{D}_{u} (\varphi) \hat{D}^{\dag}_{u}, 
\end{align}
where $\mathcal{P}_{M}(u)$ is a Gaussian probability distribution with vanishing mean and covariance matrix $M$, and $\hat{D}_{u} = e^{-i u^{T}\Omega \hat{R}/2}$ denotes the displacement operator. If $M$ is strictly positive, then $\mathcal{P}_{M}(u)$ can be written as
\begin{align}
  \mathcal{P}_{M}(u)  = \frac{e^{-1/2 u^{T}M^{-1}u}}{(2 \pi)^{n} \sqrt{\det M}}. 
\end{align}
In contrast, when $M$ has some zero eigenvalues, the definition of $\mathcal{P}_{M}(u)$ involves Dirac delta distributions. Let $K = \sum_{i=1}^{2n} \lambda_i \, \mathbf{v}_i \mathbf{v}_i^\top$
be a spectral decomposition of $K$, where $(\mathbf{v}_i)_{i=1,\dots,2n}$ forms an orthonormal set of eigenvectors. The corresponding eigenvalues satisfy
$\lambda_1, \dots, \lambda_r > 0$ and $\lambda_{r+1}, \dots, \lambda_{2n} = 0$, where $r$ denotes the rank of $K$. In this case, the distribution $P_K(\mathbf{u})$ can be expressed as
\begin{equation}
P_K(\mathbf{u}) \coloneqq
\frac{
\exp\!\left(
-\frac{1}{2}\,
\mathbf{u}^\top
\left(
\sum_{i=1}^{r} \lambda_i^{-1} \mathbf{v}_i \mathbf{v}_i^\top
\right)
\mathbf{u}
\right)
}{
(2\pi)^{r/2} \sqrt{\lambda_1 \cdots \lambda_r}
}
\prod_{i=r+1}^{2n}
\delta\!\left(\mathbf{v}_i^\top \mathbf{u}\right),
\end{equation}
where $\delta(\cdot)$ denotes the Dirac delta distribution.
It can be shown that $\mathcal{N}_K$ is a Gaussian channel that leaves the first moments invariant and transforms the covariance matrix of the input state by adding $K$, as stated in \cite{serafini2023quantum}. Let $K \in \mathbb{R}^{2n \times 2n}$ be a positive semidefinite matrix. The Gaussian noise channel $\mathcal{N}_K$ acts on the first moments and covariance matrices as
\begin{align}
    m\bigl(\mathcal{N}_K(\rho)\bigr) = m(\rho),
\quad
\Sigma\bigl(\mathcal{N}_K(\rho)\bigr) = \Sigma(\rho) + K,
\end{align}
for any $n$-mode input state $\rho$.

\subsection{Heterodyne normalisation}
The learning algorithm uses heterodyne data in the same quadrature units as the state covariance.
We therefore specify the measurement normalisation and the exact relation between its classical covariance and the quantum covariance.
These moment identities hold for non-Gaussian targets as well as Gaussian ones.
Let $|0\rangle$ denote the $n$-mode vacuum. We parametrise the heterodyne positive operator-valued measure by the physical displacement $x\in\mathbb R^{2n}$ and write it as
\begin{equation}
M(dx)=\frac{d^{2n}x}{(4\pi)^n}\,D(x)|0\rangle\langle0|D(x)^\dagger.
\label{eq:heterodyne-povm}
\end{equation}
Thus, if the conventional complex amplitude is $\alpha_j$, the real outcomes used here are $X_{2j-1}=2\operatorname{Re}\alpha_j$ and $X_{2j}=2\operatorname{Im}\alpha_j$. For every state with finite second moments, these outcomes obey
\begin{equation}
\mathbb E[X]=m,\qquad\operatorname{Cov}(X)=\Sigma+I_{2n}.
\end{equation}
For example, their characteristic function is
\begin{equation}
\mathbb E[e^{it^{\mathrm T}X}]
=e^{-\|t\|_2^2/2}\operatorname{Tr}[\rho e^{it^{\mathrm T}R}],
\qquad t\in\mathbb R^{2n},
\end{equation}
and differentiating at zero gives the two moment identities. If the measured state is Gaussian, $X$ is Gaussian with these moments. For a non-Gaussian state, the moment identities still hold, but Gaussianity of the outcome distribution must not be assumed.
\section{Proof of Theorem~\ref{theorem1}}\label{Proof of Theorem 1}

In this appendix, we prove the trace-distance bound stated in Theorem~\ref{theorem1} for the reconstruction of an approximately Gaussian mixed state. The proof is organised into three steps. First, we apply the triangle inequality to separate the non-Gaussian approximation error $\delta_{\mathrm{NG}}(\nu)$ from the error associated with estimating its moment-matched Gaussian reference state. Second, after characterising the Gaussian approximation by its covariance matrix and first moment vector, we separately bound the errors arising from estimating these two quantities.
Using the unitary invariance of the trace norm under a common displacement, we bound the covariance contribution by applying the trace norm derivative estimate for Gaussian states along the straight-line path joining $\Sigma$ and $\tilde{\Sigma}$~\cite{bittel2025optimal,bittel2025energy}. Third, we bound the displacement contribution using the Fuchs--van de Graaf inequality together with the fidelity between Gaussian states having the same covariance matrix~\cite{banchi2015quantum}. This produces a bound involving the weighted displacement norm
$\|\tilde{\Sigma}^{-1/2}(m-\tilde{m})\|_{2}$.

Throughout the proof, we use the notation introduced in the problem statement and the convention
\begin{equation}
D_{\mathrm{tr}}(\rho,\sigma)
=
\frac{1}{2}\|\rho-\sigma\|_{1}.
\end{equation}
Combining the non-Gaussian approximation, covariance estimation, and displacement estimation contributions yields Eq.~\eqref{main-trace-bound}.

The non-Gaussian approximation error is defined as
\begin{equation}
\delta_{\mathrm{NG}}(\nu)
\coloneqq
D_{\mathrm{tr}}\!\left(
\varrho_{\nu},
\rho_{\mathrm{G}}(\Sigma,m)
\right)
=
\frac{1}{2}
\left\|
\varrho_{\nu}
-
\rho_{\mathrm{G}}(\Sigma,m)
\right\|_{1}.
\label{non-Gaussian-error}
\end{equation}
Using the triangle inequality for the trace distance, we obtain
\begin{align}
D_{\mathrm{tr}}\!\left(
\varrho_{\nu},
\tilde{\rho}_{\mathrm{G}}(\tilde{\Sigma},\tilde{m})
\right)
&\leq
D_{\mathrm{tr}}\!\left(
\varrho_{\nu},
\rho_{\mathrm{G}}(\Sigma,m)
\right)
\nonumber\\
&\:+
D_{\mathrm{tr}}\!\left(
\rho_{\mathrm{G}}(\Sigma,m),
\tilde{\rho}_{\mathrm{G}}(\tilde{\Sigma},\tilde{m})
\right)
\nonumber\\
&=
\delta_{\mathrm{NG}}(\nu)
+
D_{\mathrm{tr}}\!\left(
\rho_{\mathrm{G}}(\Sigma,m),
\tilde{\rho}_{\mathrm{G}}(\tilde{\Sigma},\tilde{m})
\right).
\label{initial-decomposition}
\end{align}
Thus, the total reconstruction error separates into the error caused by replacing the non-Gaussian output with its corresponding Gaussian state and the error caused by estimating its first and second moments.

Inserting the intermediate Gaussian state $\rho_{\mathrm{G}}(\tilde{\Sigma},m)$ into the triangle inequality gives
\begin{align}
D_{\mathrm{tr}}\!\left(
\rho_{\mathrm{G}}(\Sigma,m),
\tilde{\rho}_{\mathrm{G}}(\tilde{\Sigma},\tilde{m})
\right)
&\leq
D_{\mathrm{tr}}\!\left(
\rho_{\mathrm{G}}(\Sigma,m),
\rho_{\mathrm{G}}(\tilde{\Sigma},m)
\right)
\nonumber\\
&\:+
D_{\mathrm{tr}}\!\left(
\rho_{\mathrm{G}}(\tilde{\Sigma},m),
\rho_{\mathrm{G}}(\tilde{\Sigma},\tilde{m})
\right).
\label{eq:Gaussian-error-decomposition}
\end{align}
Here, $\rho_{\mathrm{G}}(\tilde{\Sigma},m)$ is used only as an
auxiliary comparison state in the triangle inequality. The first term in
Eq.~\eqref{eq:Gaussian-error-decomposition} contains only the covariance
error, whereas the second contains only the first-moment error.

\medskip
\noindent\textit{Covariance contribution.}
The trace norm is invariant under unitary transformations. For any density matrices $\rho, \sigma$ and any unitary operator $U$
\begin{equation}
\left\|
U\rho U^{\dagger}-U\sigma U^{\dagger}
\right\|_{1}
=
\|\rho-\sigma\|_{1}.
\label{eq:unitary-invariance}
\end{equation}
A Gaussian state with first moment $m$ can be written as
\begin{equation}
\rho_{\mathrm{G}}(\Sigma,m)
=
D(m)\rho_{\mathrm{G}}(\Sigma,0)D^{\dagger}(m),
\label{eq:Gaussian-displacement}
\end{equation}
where $D(m)$ is the unitary displacement operator.
Using Eqs.~\eqref{eq:Gaussian-error-decomposition} and~\eqref{eq:Gaussian-displacement}, together with unitary invariance, we obtain
\begin{align}
\nonumber
  D_{\mathrm{tr}}\!\left(
\rho_{\mathrm{G}}(\Sigma,m),
\tilde{\rho}_{\mathrm{G}}(\tilde{\Sigma},\tilde{m})
\right)
&\leq
D_{\mathrm{tr}}\!\left(
\rho_{\mathrm{G}}(\Sigma,m),
\rho_{\mathrm{G}}(\tilde{\Sigma},m)
\right) +
D_{\mathrm{tr}}\!\left(
\rho_{\mathrm{G}}(\tilde{\Sigma},m),
\rho_{\mathrm{G}}(\tilde{\Sigma},\tilde{m})
\right),\\
  & = \frac{1}{2} \| D(m)\rho_{\mathrm{G}}(\Sigma, 0)D^\dagger(m) - D(m)\Tilde{\rho}_{\mathrm{G}}(\Tilde{\Sigma}, 0)D^\dagger(m) \|_1 + \frac{1}{2} \left\| \rho_{\mathrm{G}}(\Tilde{\Sigma}, m) - \Tilde{\rho}_{\mathrm{G}}(\Tilde{\Sigma}, \Tilde{m})  \right\|_{1}  \nonumber\\
  & = \frac{1}{2} \|\rho_{\mathrm{G}}(\Sigma, 0) - \Tilde{\rho}_{\mathrm{G}}(\Tilde{\Sigma}, 0)\|_1 + \frac{1}{2} \left\| \rho_{\mathrm{G}}(\Tilde{\Sigma}, m) - \Tilde{\rho}_{\mathrm{G}}(\Tilde{\Sigma}, \Tilde{m})  \right\|_{1}. \label{eq:zero-mean-reduction}
\end{align}

We first assume the strict uncertainty relations $\Sigma+i\Omega\succ0$ and $\tilde\Sigma+i\Omega\succ0$, so the following derivatives are taken in the interior of the physical covariance domain. For $m,\tilde m\in\mathbb R^{2n}$, the difference between the Gaussian states is given by
\begin{align}
    \rho_{\mathrm{G}}(\tilde{\Sigma}, \tilde{m}) - \rho_{\mathrm{G}}(\Sigma, m) = \int_{0}^{1} \partial_{V,v}\rho_{\mathrm{G}}\bigl(\Sigma + \alpha V, m + \alpha v\bigr) \,d\alpha, \label{2ndEq}
\end{align}
where $V \coloneqq \tilde{\Sigma} - \Sigma$ and $v \coloneqq \tilde{m} - m$.

For the covariance contribution, substituting $m=\tilde m=0$, $V=\tilde\Sigma-\Sigma$, and $v=0$ gives
\begin{align}
    \tilde{\rho}_{\mathrm{G}}(\tilde{\Sigma},0) - \rho_{\mathrm{G}}(\Sigma,0) = \int_{0}^{1} \partial_{\tilde{\Sigma} - \Sigma,0}\rho_{\mathrm{G}}\left(\Sigma + \alpha (\tilde{\Sigma} -  \Sigma), 0\right) \,d\alpha. 
\end{align}
This integral represents a continuous, straight-line path in the space of covariance matrices from $\Sigma$ (when $\alpha = 0$) to $\tilde{\Sigma}$ (when $\alpha = 1$). By the definition of the Riemann integral, the integral of a function $f(\alpha)$ on $[0,1]$ is
\begin{align}
    \int_{0}^{1} f(\alpha) \,d\alpha = \lim_{k \to \infty} \sum_{i=1}^{k} f(\alpha_i) \Delta\alpha. \label{Raimann}
\end{align}
If we break down the interval $[0,1]$ into $k$ equal steps, the step size is $\Delta\alpha  = \frac{1}{k}$. For this Riemann sum, the evaluation points $\alpha_i$ are defined as $\alpha_i = \frac{i - 1}{k}$ for $i = 1, 2,\dots, k$. Substituting $f(\alpha) = \partial_{\tilde{\Sigma} - \Sigma,0}\rho_{\mathrm{G}}\left(\Sigma + \alpha (\tilde{\Sigma} -  \Sigma), 0\right)$, we obtain
\begin{align}
    \tilde{\rho}_{\mathrm{G}}(\tilde{\Sigma},0) - \rho_{\mathrm{G}}(\Sigma,0) = \lim_{k \to \infty} \frac{1}{k} \sum_{i=1}^{k} \partial_{\tilde{\Sigma} -\Sigma,0} \rho_{\mathrm{G}}\left( \Sigma + \frac{i-1}{k}(\tilde{\Sigma} -\Sigma),0 \right).
\end{align}
Taking the trace norm of both sides and applying the triangle inequality gives
\begin{align}
    \left\| \tilde{\rho}_{\mathrm{G}}(\tilde{\Sigma},0) - \rho_{\mathrm{G}}(\Sigma,0) \right\|_1 \le \lim_{k \to \infty} \frac{1}{k} \sum_{i=1}^{k} \left\|  \partial_{\tilde{\Sigma} -\Sigma,0} \rho_{\mathrm{G}}\left( \Sigma + \frac{i-1}{k}(\tilde{\Sigma} -\Sigma),0 \right) \right\|_{1}. \label{3rdEq}
\end{align}
We use the inverse-covariance derivative estimate of Ref.~\cite{bittel2025energy}, based on the differentiability framework of Ref.~\cite{bittel2025optimal}. For a covariance $C$ in the interior of the physical domain, a real symmetric direction $X$, and a displacement direction $t$, the following bound holds (the displacement coefficient here need not be optimal)
\begin{align}
    \left\| \left. \frac{d}{d\alpha} \rho_{\mathrm{G}}(C + \alpha X, m + \alpha t) \right|_{\alpha=0} \right\|_1 \le \frac{1 + \sqrt{3}}{2} \mathrm{Tr}\left[ C^{-1}|X| \right] + 2\|C^{-1/2}t\|_2.
\end{align}
Because we are evaluating the zero-mean path ($t=0$), the second term vanishes. Applying this estimate to each term in our Riemann sum, where $C_i = \Sigma + \frac{i-1}{k}(\tilde{\Sigma} - \Sigma)$ and $X = \tilde{\Sigma} - \Sigma$, we get
\begin{align}
    \left\| \tilde{\rho}_{\mathrm{G}}(\tilde{\Sigma},0) - \rho_{\mathrm{G}}(\Sigma,0) \right\|_1  &\le \frac{1+\sqrt{3}}{2} \lim_{k\to\infty} \frac{1}{k} \sum_{i=1}^{k} \operatorname{Tr}\left[ \left( \Sigma + \frac{i-1}{k}(\tilde{\Sigma} -\Sigma) \right)^{-1} |\tilde{\Sigma} -\Sigma| \right]. \label{4thEQ}
\end{align}

Recall that the limit of the Riemann sum converges back to the definite integral
\begin{align}
    \lim_{k \to \infty} \frac{1}{k} \sum_{i=1}^{k} f\left(\frac{i-1}{k}\right) = \int_{0}^{1} f(x)\,dx .
\end{align}
Letting $x = \frac{i-1}{k}$, we rewrite the summation as an integral over the inverse covariance matrix
\begin{align}
    \left\| \tilde{\rho}_{\mathrm{G}}(\tilde{\Sigma},0) - \rho_{\mathrm{G}}(\Sigma,0) \right\|_1  &\le \frac{1+\sqrt{3}}{2} \int_{0}^{1} \operatorname{Tr}\left[ \left( \Sigma + x(\tilde{\Sigma} -\Sigma) \right)^{-1} |\tilde{\Sigma} -\Sigma| \right] \,dx. 
\end{align}
To evaluate the integral of the matrix inverse, we note that $\Sigma + x(\tilde{\Sigma} - \Sigma)$ can be written as a convex combination $(1-x)\Sigma + x\tilde{\Sigma}$. Because the matrix inverse function $f(A) = A^{-1}$ is operator convex for positive definite matrices, we have the operator inequality
\begin{align}
    \left( (1-x)\Sigma + x\tilde{\Sigma} \right)^{-1} \le (1-x)\Sigma^{-1} + x\tilde{\Sigma}^{-1}. 
\end{align}
Substituting this upper bound into the integral gives
\begin{align}
    \int_{0}^{1} \left( (1-x)\Sigma + x\tilde{\Sigma} \right)^{-1} \,dx &\le \int_{0}^{1} \left( (1-x)\Sigma^{-1} + x\tilde{\Sigma}^{-1} \right) \,dx \\
&= \left[ \left(x - \frac{x^2}{2}\right)\Sigma^{-1} + \frac{x^2}{2}\tilde{\Sigma}^{-1} \right]_{0}^{1} \\
&= \frac{1}{2}\Sigma^{-1} + \frac{1}{2}\tilde{\Sigma}^{-1} \\
&= \frac{\Sigma^{-1} + \tilde{\Sigma}^{-1}}{2}. 
\end{align}
Finally, applying this integrated upper bound to the trace operation, we arrive at the closed-form bound for the first part of the proof
\begin{align}
    \left\| \tilde{\rho}_{\mathrm{G}}(\tilde{\Sigma},0) - \rho_{\mathrm{G}}(\Sigma,0) \right\|_1  &\le \frac{1+\sqrt{3}}{2} \operatorname{Tr}\left[ \left( \frac{\Sigma^{-1} + \tilde{\Sigma}^{-1}}{2} \right) |\tilde{\Sigma} -\Sigma| \right] \\
&= \frac{1+\sqrt{3}}{4} \operatorname{Tr}\left[ \left( \Sigma^{-1} + \tilde{\Sigma}^{-1} \right) |\tilde{\Sigma} -\Sigma| \right]. \label{eq:covariance-trace-distance-bound}
\end{align}

The covariance bound also holds at the boundary of the uncertainty region. To see this, replace both covariances by $\Sigma+\tau I$ and $\tilde\Sigma+\tau I$ with $\tau>0$ and then let $\tau\downarrow0$. Their inverses converge because physical covariance matrices are positive definite. The corresponding Gaussian states converge in trace norm: adding $\tau I$ is the Gaussian displacement-noise channel in Appendix~\ref{Gaussian Noise channel}, which converges to the identity by strong continuity of displacements and dominated convergence. This justifies the limit without assuming differentiability through nonphysical directions.

\medskip
\noindent\textit{First moment contribution.}
Now, to bound the second term in Eq.~\ref{eq:zero-mean-reduction}, we consider the trace distance between the two Gaussian states $\tilde{\rho}_{\mathrm{G}}(\tilde{\Sigma},m)$ and $\tilde{\rho}_{\mathrm{G}}(\tilde{\Sigma},\tilde{m})$. For both pure and mixed Gaussian states, we apply the Fuchs--van de Graaf inequality, which upper-bounds the trace distance in terms of the Bures fidelity \cite{fuchs2002cryptographic}, $F(\rho,\sigma) = \operatorname{Tr}\!\left(\sqrt{\sqrt{\rho}\,\sigma\,\sqrt{\rho}}\right)$
with $\frac{1}{2}\|\rho - \sigma\|_1 \le \sqrt{1 - F^2(\rho,\sigma)}.$
Therefore,
\begin{align}
\frac{1}{2}\|\Tilde{\rho}_{\mathrm{G}}(\Tilde{\Sigma},m)-\Tilde{\rho}_{\mathrm{G}}(\Tilde{\Sigma},\tilde{m})\|_1
\le \sqrt{1 - F^2\!\left(\Tilde{\rho}_{\mathrm{G}}(\Tilde{\Sigma},m),\Tilde{\rho}_{\mathrm{G}}(\Tilde{\Sigma},\tilde{m})\right)}. \label{Eq:FidInequality}
\end{align}
For two Gaussian states sharing the same covariance matrix $\Tilde{\Sigma}$ but differing in displacement vectors $m$ and $\tilde{m}$, the fidelity~\cite{banchi2015quantum} is given by  $F^2
= \exp\!\left(-\frac{1}{4}(m-\tilde{m})^\top \Tilde{\Sigma}^{-1}(m-\tilde{m})\right)$.
Here $F$ is the root fidelity, so $F^2$ is the squared fidelity.
With $[R_j,R_k]=2i\Omega_{jk}$ and vacuum covariance $I_{2n}$, the displacement exponent is $-\tfrac14(m-\tilde m)^{\mathrm T}\tilde\Sigma^{-1}(m-\tilde m)$.
This agrees with the coherent-state overlap when $\tilde\Sigma=I_{2n}$.

Substituting this into Eq.~\eqref{Eq:FidInequality} yields
\begin{align}
\frac{1}{2}\|\Tilde{\rho}_{\mathrm{G}}(\Tilde{\Sigma},m)-\Tilde{\rho}_{\mathrm{G}}(\Tilde{\Sigma},\tilde{m})\|_1
\le
\sqrt{
1-\exp\!\left(
-\frac{1}{4}(m-\tilde{m})^\top \Tilde{\Sigma}^{-1}(m-\tilde{m})
\right)
}.
\end{align}
To simplify this expression, we use the elementary bound $e^{-x}\geq 1-x$ for all real $x$, which implies $1-e^{-x}\leq x.$ 
Setting $x=\frac{1}{4}(m-\tilde{m})^\top \Tilde{\Sigma}^{-1}(m-\tilde{m}),$
we obtain
\begin{align}
\frac{1}{2}\|\Tilde{\rho}_{\mathrm{G}}(\Tilde{\Sigma},m)-\Tilde{\rho}_{\mathrm{G}}(\Tilde{\Sigma},\tilde{m})\|_1
\le
\sqrt{
\frac{1}{4}(m-\tilde{m})^\top \Tilde{\Sigma}^{-1}(m-\tilde{m})
}.
\end{align}

Since $\Tilde{\Sigma}$ is positive definite, its inverse admits the symmetric factorization $\Tilde{\Sigma}^{-1}
=
\Tilde{\Sigma}^{-1/2}\Tilde{\Sigma}^{-1/2}.$
Thus, the quadratic form can be rewritten exactly as
\begin{align}
(m-\tilde{m})^\top \Tilde{\Sigma}^{-1}(m-\tilde{m})
&=
(m-\tilde{m})^\top
\Tilde{\Sigma}^{-1/2}
\Tilde{\Sigma}^{-1/2}
(m-\tilde{m})
\nonumber\\
&=
\left(
\Tilde{\Sigma}^{-1/2}(m-\tilde{m})
\right)^\top
\left(
\Tilde{\Sigma}^{-1/2}(m-\tilde{m})
\right)
\nonumber\\
&=
\left\|
\Tilde{\Sigma}^{-1/2}(m-\tilde{m})
\right\|_2^2.
\end{align}

Substituting this identity into the bound gives
\begin{align}
\frac{1}{2}\|\Tilde{\rho}_{\mathrm{G}}(\Tilde{\Sigma},m)-\Tilde{\rho}_{\mathrm{G}}(\Tilde{\Sigma},\tilde{m})\|_1
&\le
\sqrt{
\frac{1}{4}
\left\|
\Tilde{\Sigma}^{-1/2}(m-\tilde{m})
\right\|_2^2
}
\nonumber\\
&=
\frac{1}{2}
\left\|
\Tilde{\Sigma}^{-1/2}(m-\tilde{m})
\right\|_2. \label{eq:first-moment-trace-distance-bound}
\end{align}

Finally, substituting Eqs.~\eqref{eq:covariance-trace-distance-bound}
and \eqref{eq:first-moment-trace-distance-bound} into
Eqs.~\eqref{eq:Gaussian-error-decomposition} and
\eqref{initial-decomposition} gives
\begin{align}
&D_{\mathrm{tr}}\!\left(
\varrho_{\nu},
\tilde{\rho}_{\mathrm{G}}(\tilde{\Sigma},\tilde{m})
\right)
\nonumber\\
&\leq
\delta_{\mathrm{NG}}(\nu)
+
\frac{1+\sqrt{3}}{8}
\operatorname{Tr}\!\left[
\left(\Sigma^{-1}+\tilde{\Sigma}^{-1}\right)
\left|\tilde{\Sigma}-\Sigma\right|
\right]
\nonumber\\
&\:+
\frac{1}{2}
\left\|
\tilde{\Sigma}^{-1/2}(m-\tilde{m})
\right\|_{2}.
\label{eq:final-trace-distance-bound}
\end{align}
This proves the trace-distance form of Theorem~\ref{theorem1}.

Equivalently, multiplying Eq.~\eqref{eq:final-trace-distance-bound} by
$2$ gives the trace-norm form
\begin{align}
&
\left\|
\varrho_{\nu}
-
\tilde{\rho}_{\mathrm{G}}(\tilde{\Sigma},\tilde{m})
\right\|_{1}
\nonumber\\
&\leq
2\delta_{\mathrm{NG}}(\nu)
+
\frac{1+\sqrt{3}}{4}
\operatorname{Tr}\!\left[
\left(\Sigma^{-1}+\tilde{\Sigma}^{-1}\right)
\left|\tilde{\Sigma}-\Sigma\right|
\right]
\nonumber\\
&\:+
\left\|
\tilde{\Sigma}^{-1/2}(m-\tilde{m})
\right\|_{2}.
\end{align}


\section{Proof of Theorem~\ref{theorem2}}\label{Proof of Theorem2}

The heterodyne outcomes in this proof need not be Gaussian. We use only Eq.~\eqref{eq:heterodyne-subgaussian}, independence between copies, and the exact identity $H=\Sigma+I_{2n}$. We first prove a concentration estimate with explicit constants, then establish physicality and trace-distance recovery, and finally verify the statistical assumption for stochastic phase noise.

\subsection{Concentration of the empirical moments}
We first control the empirical mean and covariance after whitening the measurement distribution.
The argument uses only the stated sub-Gaussian condition, and explicitly accounts for centring the covariance with the empirical mean.
This yields one event on which both moment errors are controlled.

\begin{lemma}(Centred empirical covariance of sub-Gaussian vectors.)
\label{lem:subgaussian-moments}
Let $Z_1,\ldots,Z_N\in\mathbb R^d$ be independent copies of a centred random vector with $\mathbb E[ZZ^{\mathrm T}]=I_d$ and
\begin{equation}
\mathbb E[e^{u^{\mathrm T}Z}]\leq
\exp(\kappa^2\|u\|_2^2/2),\qquad u\in\mathbb R^d,
\label{eq:isotropic-mgf}
\end{equation}
where $\kappa\geq1$. Write $\overline Z=N^{-1}\sum_jZ_j$, let
$q=d\ln9+\ln(4/\delta)$, and define
$S_Z=N^{-1}\sum_j(Z_j-\overline Z)(Z_j-\overline Z)^{\mathrm T}$.
Then, with probability at least $1-\delta$,
\begin{align}
\|\overline Z\|_2&\leq2\kappa\sqrt{q/N},\label{eq:mean-concentration}\\
\|S_Z-I_d\|_\infty&\leq24\kappa^2\left(\sqrt{q/N}+q/N\right).
\label{eq:covariance-concentration}
\end{align}
\end{lemma}
\noindent\textit{Proof.}
We use the standard finite-net argument for sub-Gaussian covariance concentration~\cite{vershynin2018high,vershynin2012introduction}, spelling out constants and empirical centring. Choose a $1/4$-net $\mathcal N$ of the Euclidean unit sphere with $|\mathcal N|\leq9^d$. The net inequalities are
\begin{equation}
\|v\|_2\leq\frac43\max_{u\in\mathcal N}|u^{\mathrm T}v|,
\qquad
\|A\|_\infty\leq2\max_{u\in\mathcal N}|u^{\mathrm T}Au|
\label{eq:net-inequalities}
\end{equation}
for any real vector $v$ and real symmetric matrix $A$. For the second inequality, choose a unit vector maximising $|v^{\mathrm T}Av|$ and a net point at distance at most $1/4$; their quadratic forms differ by at most $\|A\|_\infty/2$. The first inequality follows in the same way from a linear form. The net cardinality follows by comparing volumes of disjoint balls of radius $1/8$ centred at a maximal separated set on the sphere.

For each unit vector $u$, Eq.~\eqref{eq:isotropic-mgf} and Chernoff's bound give
\begin{equation}
\Pr\!\left[|u^{\mathrm T}\overline Z|>
\kappa\sqrt{2q/N}\right]\leq2e^{-q}.
\end{equation}
A union bound over $\mathcal N$ has failure probability at most $2\cdot9^de^{-q}=\delta/2$ and gives
$\|\overline Z\|_2\leq(4\sqrt2/3)\kappa\sqrt{q/N}\leq2\kappa\sqrt{q/N}$.

For the uncentred second moment, fix a unit vector $u$ and write $Y=u^{\mathrm T}Z$ and $W=Y^2-1$. Then $\mathbb E[Y^2]=1$, $\mathbb E[W]=0$, and
$\Pr(|Y|>s)\leq2e^{-s^2/(2\kappa^2)}$. Integration of this tail gives, for every integer $p\geq2$,
\begin{equation}
\mathbb E|Y|^{2p}\leq2(2\kappa^2)^p p!,\qquad
\mathbb E|W|^p\leq2^p\mathbb E|Y|^{2p}
\leq2(4\kappa^2)^p p!.
\label{eq:scalar-moment-bound}
\end{equation}
The middle inequality uses $\mathbb E|Y|^{2p}\geq(\mathbb E Y^2)^p=1$. Thus, with $v^2=64\kappa^4$ and $b_0=4\kappa^2$,
$\mathbb E|W|^p\leq(p!/2)v^2b_0^{p-2}$. Expanding the exponential series and using $\mathbb E W=0$ gives
\begin{equation}
\log\mathbb E[e^{\lambda W}]
\leq\frac{v^2\lambda^2}{2(1-b_0|\lambda|)},
\qquad |\lambda|<b_0^{-1}.
\label{eq:scalar-bernstein-mgf}
\end{equation}
Applying Chernoff's bound to a sum of $N$ independent copies, for either sign of $W$, yields
\begin{equation}
\Pr\!\left[
\left|\frac1N\sum_{j=1}^N W_j\right|
>8\sqrt2\,\kappa^2\sqrt{q/N}+4\kappa^2q/N
\right]\leq2e^{-q}.
\label{eq:scalar-bernstein-tail}
\end{equation}
For completeness, the upper-tail estimate follows from Eq.~\eqref{eq:scalar-bernstein-mgf} with
$\lambda=s/(1+b_0s)$, where $s=\sqrt{2q/(Nv^2)}$, and threshold
$\sqrt{2Nv^2q}+b_0q$ for the sum. The lower tail follows by replacing $W$ by $-W$.

A second union bound over $\mathcal N$, again with failure probability at most $\delta/2$, and Eq.~\eqref{eq:net-inequalities} imply
\begin{equation}
\left\|\frac1N\sum_{j=1}^NZ_jZ_j^{\mathrm T}-I_d\right\|_\infty
\leq16\sqrt2\,\kappa^2\sqrt{q/N}+8\kappa^2q/N.
\label{eq:uncentred-moment-bound}
\end{equation}
Finally, empirical centring gives the exact identity
\begin{equation}
S_Z=\frac1N\sum_{j=1}^NZ_jZ_j^{\mathrm T}
-\overline Z\,\overline Z^{\mathrm T}.
\label{eq:exact-centring}
\end{equation}
On the intersection of the two events, the additional term has norm at most $4\kappa^2q/N$. Since $16\sqrt2<24$ and $12<24$, Eqs.~\eqref{eq:uncentred-moment-bound} and \eqref{eq:exact-centring} prove Eq.~\eqref{eq:covariance-concentration}. The two failure probabilities add to $\delta$.\hfill$\square$

\subsection{Physical reconstruction and the sample bound}
We now transfer the concentration event to the quantum reconstruction.
Inflating the empirical heterodyne covariance ensures that the candidate satisfies the uncertainty relation on that event.
The deterministic bound then converts the moment estimates into the required trace-distance guarantee and an explicit sample threshold.

Set $d=2n$ and $Z_j=H^{-1/2}(X_j-m)$. These vectors satisfy Lemma~\ref{lem:subgaussian-moments}. In particular,
\begin{equation}
H^{-1/2}\widehat H H^{-1/2}=S_Z,
\qquad H^{-1/2}(\widehat m_H-m)=\overline Z.
\label{eq:whitened-identities}
\end{equation}
On an event $\mathcal E$ of probability at least $1-\delta$,
\begin{equation}
(1-\zeta_N)H\preceq\widehat H\preceq(1+\zeta_N)H,
\qquad
\|H^{-1/2}(\widehat m_H-m)\|_2\leq2\kappa\sqrt{q_\delta/N}.
\label{eq:joint-moment-event}
\end{equation}
If $\zeta_N\leq1/2$, the candidate in Eq.~\eqref{eq:physical-estimator} consequently satisfies
\begin{equation}
0\preceq\widetilde\Sigma-\Sigma
\preceq\frac{2\zeta_N}{1-\zeta_N}(\Sigma+I_{2n}).
\label{eq:inflated-covariance-error}
\end{equation}
Since $\widetilde\Sigma+i\Omega\succeq\Sigma+i\Omega\succeq0$, this candidate is physical on $\mathcal E$ and no fallback is used there. Notice that physicality requires the uncertainty relation, not just positivity of the real covariance matrix.

By inversion monotonicity, $\widetilde\Sigma^{-1}\preceq\Sigma^{-1}$. Since the error in Eq.~\eqref{eq:inflated-covariance-error} is positive semidefinite, the covariance contribution to Theorem~\ref{theorem1} is bounded by
\begin{align}
&\frac{1+\sqrt3}{8}\operatorname{Tr}\!\left[
(\Sigma^{-1}+\widetilde\Sigma^{-1})|\widetilde\Sigma-\Sigma|\right]\nonumber\\
&\quad\leq\frac{1+\sqrt3}{2}
\frac{\zeta_N}{1-\zeta_N}\operatorname{Tr}[\Sigma^{-1}(\Sigma+I_{2n})]
=\frac{1+\sqrt3}{2}\frac{\zeta_N}{1-\zeta_N}E.
\label{eq:statistical-covariance-term}
\end{align}
The trace inequalities hold without a commutativity assumption: multiplying a positive semidefinite inequality by a positive semidefinite weight and taking its trace preserves the inequality.

For the first moments, use $H=\Sigma+I_{2n}$ to obtain
\begin{align}
\|\widetilde\Sigma^{-1/2}(\widehat m_H-m)\|_2
&\leq\|\Sigma^{-1/2}(\widehat m_H-m)\|_2\nonumber\\
&\leq\sqrt{1+\|\Sigma^{-1}\|_\infty}\,
\|H^{-1/2}(\widehat m_H-m)\|_2.
\label{eq:correct-mean-weight}
\end{align}
Here $\|\Sigma^{-1/2}H^{1/2}\|_\infty^2=1+\|\Sigma^{-1}\|_\infty$; no assumption $\Sigma\succeq I_{2n}$ is made. Combining Eqs.~\eqref{eq:joint-moment-event}, \eqref{eq:statistical-covariance-term}, and \eqref{eq:correct-mean-weight} with Theorem~\ref{theorem1} proves Eq.~\eqref{eq:finite-N-trace-bound}.

It remains to justify the finite-sample threshold. Let $\varepsilon_s=\varepsilon_{\mathrm{tot}}-U\in(0,1]$ and $t=q_\delta/N$. Suppose
\begin{equation}
N\geq2^{16}\kappa^4q_\delta B^2/\varepsilon_s^2.
\label{eq:explicit-sample-threshold}
\end{equation}
Then $\sqrt t\leq\varepsilon_s/(256\kappa^2B)$. Since $B\geq E\geq2$, we have $t\leq1$ and
$\zeta_N\leq48\kappa^2\sqrt t\leq3/32<1/2$. Also,
$1+\|\Sigma^{-1}\|_\infty\leq E$. Therefore, on $\mathcal E$,
\begin{align}
D_{\mathrm{tr}}(\varrho_\nu,\widetilde\rho_{\mathrm G})
&\leq U+\bigl[48(1+\sqrt3)+1\bigr]\kappa^2E\sqrt t\nonumber\\
&\leq U+\frac{48(1+\sqrt3)+1}{256}\,\varepsilon_s
\leq\varepsilon_{\mathrm{tot}}.
\label{eq:final-statistical-guarantee}
\end{align}
In the first line we used $(1-\zeta_N)^{-1}\leq2$ and
$\kappa\sqrt E\leq\kappa^2E$. This proves Theorem~\ref{theorem2}, with the stated universal constant. The only approximation error retained throughout is $\delta_{\mathrm{NG}}(\nu)\leq U$; it does not vanish with $N$.

\subsection{Proof of Proposition~\ref{prop:phase-subgaussian}}
Finally, we verify the sub-Gaussian promise for the phase-noise family.
Although the unconditional data are generally non-Gaussian, their conditional covariances are uniformly bounded and their conditional means range over a bounded set.
Combining these two controls gives the advertised tail parameter and inverse-covariance bound.
Conditioned on the phases, the heterodyne outcome is Gaussian with
\begin{equation}
m_\phi=O_\phi m_{\mathrm{in}},\qquad
H_\phi=O_\phi\Sigma_{\mathrm{in}}O_\phi^{\mathrm T}+I_{2n}.
\end{equation}
The law of total covariance gives
\begin{equation}
m=\mathbb E_\phi[m_\phi],\qquad
H=\mathbb E_\phi[H_\phi]+\operatorname{Cov}_\phi(m_\phi)
\succeq(1+a)I_{2n}.
\end{equation}
For any $u$, orthogonality implies that the random variable $u^{\mathrm T}m_\phi$ lies in an interval of length at most $2\|u\|_2\|m_{\mathrm{in}}\|_2$. The bounded-variable exponential estimate (Hoeffding's lemma) gives
\begin{equation}
\mathbb E_\phi[e^{u^{\mathrm T}(m_\phi-m)}]
\leq\exp\!\left(\frac{\|m_{\mathrm{in}}\|_2^2\|u\|_2^2}{2}\right).
\end{equation}
One quick justification is to differentiate the centred log moment-generating function twice: its second derivative is a variance under an exponentially tilted measure on the same interval, and is at most one quarter of the squared interval length. Integrating twice from zero proves the estimate.

Since $H_\phi\preceq(1+b)I_{2n}$, averaging the conditional Gaussian moment-generating functions now gives
\begin{align}
\mathbb E[e^{u^{\mathrm T}(X-m)}]
&=\mathbb E_\phi\!\left[e^{u^{\mathrm T}(m_\phi-m)+u^{\mathrm T}H_\phi u/2}\right]\nonumber\\
&\leq\exp\!\left(\frac{1+b+\|m_{\mathrm{in}}\|_2^2}{2}\|u\|_2^2\right)
\leq\exp\!\left(\frac{\kappa^2}{2}u^{\mathrm T}Hu\right),
\end{align}
where $\kappa$ is defined in Eq.~\eqref{eq:phase-kappa}. Finally,
$\Sigma=\mathbb E_\phi[O_\phi\Sigma_{\mathrm{in}}O_\phi^{\mathrm T}]+\operatorname{Cov}_\phi(m_\phi)\succeq aI_{2n}$, which gives $E\leq2n(1+a^{-1})$. This completes the proof.


\section{Unitary averaging in a nutshell}
\label{Unitary Averaging in a nutshell}
We review the encoding, phase noise, and vacuum heralding that define the UA example.
The aim is to identify the exact conditional ensemble entering the reconstruction guarantee, including the heralding weights and phase conventions.
The number of parallel channel uses in one attempt is kept distinct from the number of accepted tomography samples.

\begin{figure}[!htb]
	\centering
		 \includegraphics[width=0.6\linewidth]{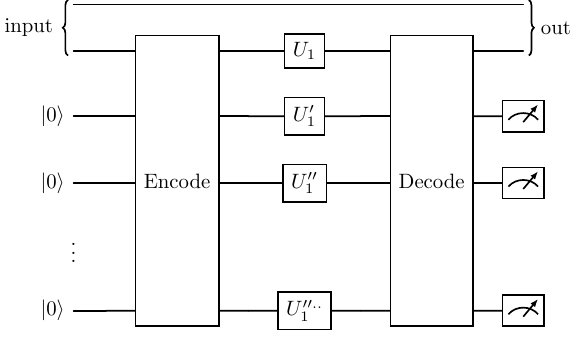}
		\caption{   A passive unitary averaging scheme utilises a beam splitter network for redundant encoding over $n_\mathrm{UA}$ modes. One mode of a two-mode state is evenly distributed across $n_\mathrm{UA}$ transmission modes. Each transmitted mode undergoes an independent phase shift. The decoding network reverses the encoding process, and upon heralding $(n_\mathrm{UA})-1$ error modes in the vacuum state, a two-mode output is retained. The figure depicts the scheme of Ref.~\cite{swain2024improving}.
  \label{CVUAcir}}
\end{figure}

We begin by reviewing previous work on Gaussian systems, in which the impact of unitary averaging (UA) on a continuous-variable (CV) system is modelled using a two-mode squeezed vacuum state, as described in \cite{swain2024improving}.
In this model, a passive beam splitter encoding network is employed to place the input state into an equal superposition of $n_\mathrm{UA}$ spatial modes. Each transmitted mode undergoes an independent phase shift $U_j=e^{i\phi_j a_j^\dagger a_j}$. The analysis here concerns these phase shifts, not arbitrary single-mode Gaussian unitaries. The decoding network then inverts the encoding such that each superposition should constructively interfere in the retained signal mode. The remaining `error modes' are then heralded in the vacuum state, leaving the conditional two-mode state described below. The behaviour is characterised by the resulting output squeezing, purity, and entanglement. 
The two-mode squeezed vacuum provides an entangled probe for characterising this heralded channel. 
Figure~\ref{CVUAcir} illustrates the circuit discussed in Ref. \cite{swain2024improving}. Below, we provide a brief summary of the corresponding results.

For a fixed phase realisation, the normalised output is
    \begin{align}
    \ket{\psi_{\text{out}}} &= \frac{1}{\mathbf{N}} \left(\cosh{r}\right)^{-1}\sum_{M=0}^{\infty} (-1)^{M} [e^{i \phi_{\beta}} \tanh{r'} ]^{M} \ket{M, M}\nonumber\\
    &=\hat{S}\left(\chi'\right)\ket{0}\ket{0}, 
\end{align}
where $\chi'=r'e^{i\phi_{\beta}}$ and the phase terms $\phi_j$ vary randomly and independently around some mean value. We define 
    \begin{align}
\alpha e^{i\phi_{\beta}}= \frac{e^{ i \phi_{1}} + e^{ i \phi_{2}} +\dots+e^{ i \phi_{n_{\text{UA}}}}}{n_\text{UA}} 
\end{align}
and $\tanh{r'} = \alpha\tanh{r}$ with normalisation constant 
\begin{equation}
\mathbf{N}  = \frac{\cosh{r'}}{\cosh{r}}
\end{equation} 
and 
probability of success  $P = |\mathbf{N}|^{2}$.
Here \begin{equation}
\hat S(\chi)=\exp(\chi^*a_1a_2-\chi a_1^\dagger a_2^\dagger), 
\end{equation}
fixing the phase convention in the state expansion. The normalisation $\mathbf N$ is not the tomography sample size $N$.
The ensemble-averaged covariance matrix is
\begin{equation}
     \Sigma_{\text{out}(4 \cross 4)} = \mathbb E_{\mathrm{acc}}\left[\begin{pmatrix}
          A  &  C  \\
          C^{T} &  B
     \end{pmatrix}\right], \label{eq:covariance matrixTMSV}
\end{equation}
where
\begin{align}
    A = B 
    & = \begin{pmatrix}
        \frac{1+ \tanh{r^{'}}^{2}}{1 - \tanh{r^{'}}^{2}} & 0\\
        0 &  \frac{1+ \tanh{r^{'}}^{2}}{1 - \tanh{r^{'}}^{2}}
    \end{pmatrix}, \label{aeq}
\end{align}
and
\begin{align}
    C   
    & = \begin{pmatrix}
       - \frac{2\tanh{r^{'}}}{1 - \tanh{r^{'}}^{2}}\cos\left(\phi_{\beta}\right) & \mathscr{C}\\
 \mathscr{C} & \frac{2\tanh{r^{'}}}{1 - \tanh{r^{'}}^{2}}\cos\left(\phi_{\beta}\right) 
    \end{pmatrix}, \label{ceq}
\end{align}

with
\begin{equation}
\mathscr{C} = -\frac{2\tanh r'}{1-\tanh^2r'}\sin\phi_\beta. \label{curly c}
\end{equation}
The correlation block satisfies $C=C^{T}$. 

Here $n_{\mathrm{UA}}$ counts the parallel noisy channel uses in one protocol attempt, not the number of repeated attempts. All phase expectations in this appendix are taken with respect to the distribution conditioned on successful heralding. If the prior phase distribution is invariant under $\boldsymbol\phi\mapsto-\boldsymbol\phi$, then so is the accepted distribution: the success probability is unchanged by this transformation. The sine term is odd, so $\mathbb E_{\mathrm{acc}}[\mathscr C]=0$ exactly. This gives the covariance matrix used in Eq.~\eqref{eq:UA-true-covariance}. No small-noise factorisation is used in any certified statement.
In summary, this appendix reviews the UA construction of Ref.~\cite{swain2024improving}. The noise-reduction benchmarks in that reference are prior results, not numerical results established in the present paper. Here we use the exact heralded ensemble to derive a Gaussian-reconstruction guarantee and to account for its sampling and heralding costs.

\end{widetext}

\bibliography{refMain}

\end{document}